\documentclass[preprint,12pt]{elsarticle}
\journal{Journal of High Energy Astrophysics}
\usepackage{amsmath,amssymb}
\usepackage{graphicx}
\usepackage{xcolor}      
\usepackage{hyperref} 
\usepackage{lineno}
\usepackage{geometry}
\usepackage{tabularx}
\usepackage{array}

\begin{document}

\begin{frontmatter}

\title{\textbf{Constraining Exponential \( f(Q) \) Gravity with Cosmic Chronometers and Supernovae: A Data-Driven Analysis}}

\author[amity]{Sanjeeda Sultana}
\ead{sanjeeda.sultana0401@gmail.com; sanjeeda.sultana1@s.amity.edu}
\author[amity]{Surajit Chattopadhyay\corref{cor1}}
\cortext[cor1]{Corresponding author}
\ead{schattopadhyay1@kol.amity.edu; surajitchatto@outlook.com}

\address[amity]{Department of Mathematics, Amity University Kolkata, Major Arterial Road, Action Area II, Rajarhat, Newtown, Kolkata 700135, India}

\begin{abstract}
The current paper reports an investigation of the cosmological implications of symmetric teleparallel gravity within a modified $f(Q)$ theory. We construct a specific exponential $f(Q)$ model as $f(Q) = Q + \eta_1 Q_0\left(1 - e^{-\eta_2 \sqrt{Q/Q_0}}\right)$, designed to smoothly deviate from General Relativity and accommodate both early-time inflation and late-time accelerated expansion. By employing Markov Chain Monte Carlo (MCMC) methods, we constrain the model parameters $\eta_1$, $\eta_2$, $H_0$, and $\Omega_{m_0}$ using a combination of cosmic chronometers (CC), Pantheon, and Pantheon$^+$ Supernovae datasets. Our analysis demonstrates that the model consistently supports a late-time acceleration scenario and is in good agreement with current cosmological observations. We extensively analyze the dynamical behavior of the model using key cosmological diagnostics, including the deceleration parameter, equation of state, energy density parameters, Statefinder, and Om diagnostics. The reconstructed Hubble parameter $H(z)$ and distance modulus $\mu(z)$ show strong consistency with $\Lambda$CDM and observational data, while subtle deviations at higher redshifts highlight the value of multi-probe observations. In addition, the examination of energy conditions shows that, in accordance with cosmic acceleration, the Strong Energy Condition (SEC) is broken at lower redshifts while the Dominant Energy Condition (DEC) and Null Energy Condition (NEC) are satisfied. Cosmic age estimates from the model are consistently in agreement with Planck constraints. \textcolor{black}{Our results indicate the viability of exponential $f(Q)$ gravity providing a consistent framework for exploring cosmic evolution and highlighting the significance of multi-probe cosmological measurements for further developments. A comparative statistical analysis reveals that while $\Lambda$CDM remains statistically preferred based on AIC and BIC criteria, the exponential $f(Q)$ model yields comparable fits and remains a theoretically motivated and viable alternative for describing cosmic acceleration.}
\end{abstract}


\end{frontmatter}
\tableofcontents
\section{Introduction}
\label{sec:intro}
The finding that the universe is expanding with acceleration, contrary to what was believed earlier, resulted in a significant change in modern cosmology \cite{Weinberg2013, Filippenko2004}.  This phenomenon was first discovered through analysis of observations of Type Ia Supernovae \cite{Perlmutter2003, Riess2001}, which appeared fainter than expected in a universe that was slowing down. Later confirmations from measurements of the Cosmic Microwave Background (CMB) and Baryon Acoustic Oscillations (BAO) strengthened the case for this accelerated expansion \cite{Eisenstein2005, Anderson2014, Alam2017}.  Dark energy, an exotic component that accounts for the majority of the cosmic energy budget, is thought to be driving the universe's late-time acceleration. One of the most pressing issues in theoretical and observational cosmology is still determining what dark energy is \cite{Copeland2006,Frieman2008,Joyce2016}. 

The cosmological constant \(\Lambda\), which is defined by a constant equation of state (EoS) parameter \(w = -1\), is still the most straightforward and commonly accepted candidate for dark energy, but it faces many theoretical obstacles, including the coincidence and fine-tuning issues \cite{Padmanabhan2003}. Several phenomenological dark energy models with time-varying EoS parameters have been developed in response to these problems, enabling a dynamical evolution of the dark energy component \cite{Chevallier2001, Linder2003}. These models provide a more comprehensive framework for analyzing current and future observational data and attempt to capture potential departures from the typical behavior of the \(\Lambda\) CDM model. There are two basic categories into which phenomenological dark energy models can be generally divided. Field theory serves as the motivation for scalar field models like quintessence and k-essence, which describe dark energy using a dynamical scalar field that changes over time \cite{Caldwell1998, ArmendarizPicon2000}. However, a number of models that are not directly based on basic field theory have also drawn interest. These include the holographic dark energy (HDE) models, which are motivated by the holographic principle and constrain the dark energy density using an infrared cutoff \cite{Li2004}, and the Chaplygin gas models, which unify dark energy and dark matter through a single fluid with an exotic equation of state \cite{Kamenshchik2001}. Because of their distinct physical motivations and cosmological implications, the Chaplygin gas and holographic dark energy (HDE) models are two of the most prominent non-field-theoretic phenomenological models. With its exotic equation of state \( p = -A/\rho \), where \(A\) is a positive constant, the Chaplygin gas model was first put forth as a unification scheme for dark matter and dark energy \cite{Kamenshchik2001, Bento2002}. To make this model more consistent with observational data, several authors have reported various generalizations. The holographic principle, on the other hand, which holds that the area of a space's boundary determines the number of degrees of freedom in that space, serves as the inspiration for the HDE model. On the other hand, the HDE model is inspired by the holographic principle \cite{Li2004}, which proposes that the number of degrees of freedom in a volume of space is proportional to the area of its boundary. Instead of considering a distinct dark energy component to explain the observed cosmic acceleration, the modified theories of gravity attribute it to changes in the gravitational sector. As the late-time acceleration results from the geometry of spacetime itself, this method is frequently referred to as \textit{geometric dark energy} \cite{Capozziello2006, Sotiriou2010}. A key feature of modified gravity is that it does not rely on a particular form of dark energy fluid or potential, thereby providing a more fundamental description of the acceleration mechanism \cite{Nojiri2011}. These models extend general relativity through functions of curvature, torsion, or non-metricity scalars and are constrained by both cosmological observations and local gravity tests. Comprehensive reviews on modified theories of gravity, including \(f(R)\), \(f(T)\), and \(f(Q)\) frameworks, can be found in \cite{Sotiriou2010, Nojiri2011, Bahamonde2022}, offering detailed discussions on theoretical foundations and observational constraints. 

In the arena of modified gravity, \( f(Q) \) gravity \cite{narawade2025insights,koussour2022cosmic} has emerged as a compelling alternative to Einstein's General Relativity, with the focus on resolving cosmological puzzles such as dark energy and the accelerated expansion of the universe without resorting to exotic forms of matter.  Unlike metric-based modifications of gravity, \( f(Q) \) gravity \cite{dimakis2025static} is developed within the framework of symmetric teleparallel gravity, wherein the gravitational interaction arises from a non-vanishing non-metricity scalar \( Q \), rather than from curvature or torsion. This distinctive geometric formulation interprets gravity as an effect of the ``stretching'' or ``contraction'' of spacetime, in contrast to the conventional notion of ``bending.'' This theory involves a set of field equations by generalizing the teleparallel equivalent of GR, where the gravitational Lagrangian is linearly proportional to $Q$ and consequently, to an arbitrary function $ f(Q)$ \cite{Lymperis2022,Paul2022,Narawade2023a}. These changes create additional degrees of freedom and provide a rich environment for investigating astrophysical and cosmological phenomena outside of the conventional paradigm.

In recent years, $f(Q)$ gravity has emerged as a widely studied alternative theory of gravity, particularly in the field of cosmology \cite{Lymperis2022,Paul2022,Narawade2023a,Narawade2023b,Dimakis2023}. Researchers have applied this framework to several areas, including the evolution of large-scale cosmic structures\cite{Sokoliuk2023}, relativistic formulations of Modified Newtonian Dynamics (MOND) \cite{Milgrom2019,DAmbrosio2020}, models of a non-singular bouncing universe \cite{Bajardi_2020_135,Agrawal2021,Gadbail2023}, and scenarios in quantum cosmology \cite{Dimakis_2021_38,Bajardi:2023fQ}. Significant attention has also been directed toward testing and placing observational constraints on $f(Q)$ gravity theories \cite{Dialektopoulos_2019_79,Ayuso_2021_103,Barros_2020_30,koussour2023thermodynamical,koussour2023observational}. Proposed extensions to the standard formulation include the incorporation of boundary terms \cite{capozziello2023role,de2024non,paliathanasis2024minisuperspace,bhagat2024observational,maurya2025implication,samaddar2024holographic,shekh2023new,myrzakulov2024observational} and scalar fields \cite{jarv2018nonmetricity,Harko_2018_98} that couple non-minimally to the geometry. Numerous investigations have explored the role of $f(Q)$ gravity \cite{banerjee2021wormhole,javed2023thermal,chanda2022evolution} in describing black hole solutions, alternative stellar configurations \cite{chanda2022evolution,maurya2022exploring,bhar2023physical}, and wormhole geometries \cite{banerjee2021wormhole,javed2023thermal}. These efforts underscore the theory's rich structure and its potential for testability. Notably, studying stellar systems that deviate from the predictions of general relativity could provide valuable observational insights for constraining models within the $f(Q)$ gravity framework \cite{d2022forecasting}. The authors \cite{Bhagat_2023_41,goswami2024flrw,Bhagat_2023_42} propose to develop a cosmological model of the universe based on Weyl type \( f(Q) \) gravity. This model has a notable feature: the universe was decelerating in the past, but after a certain epoch, it began to accelerate and continues to do so in the present. Theories of gravity, including $f(Q)$ gravity and teleparallel gravity in general, have presented new challenges. One specific example is the Hamiltonian analysis of $f(Q)$ gravity, which must address certain technical difficulties that may necessitate the development of new techniques \cite{d2020adm,hu2022adm}. This work focuses on constraining the cosmological and model parameters using different combined datasets for the adopted $f(Q)$ gravity model. We have adopted a Bayesian statistical analysis using MCMC simulation with the CC \cite{shukla2024cosmic} and the latest extensive Pantheon datasets \cite{shukla2024cosmic,arora2022bulk,zhadyranova2024constraints} to constrain these functions. The results show the chosen $f(Q)$ gravity model as a viable candidate to describe the late-time acceleration of the universe.

\textcolor{black}{Although the current study has significant motivation from the work of \cite{mhamdi2024constraints}, the study reported here, adopts a methodology distinct from the mentioned one in the sense that here by proposing a specific exponential functional form of $f(Q)$ gravity and employing MCMC techniques we attempt to constrain the model parameters using cosmic chronometers (CC), Pantheon, and Pantheon$^+$ Supernovae datasets, contrary to the earlier study \cite{mhamdi2024constraints}, which utilized alternative functional forms of $f(Q)$ and focused on parameter estimation using combinations such as Pantheon+H(z) and Pantheon+H(z)+RSD. In addition to parameter constraints, the current analysis offers a comprehensive dynamical investigation, including the evolution of the equation of state parameter and the deceleration parameter, Statefinder and $Om(z)$ diagnostics, as well as a discussion on energy conditions and cosmological age estimates. In contrast, previous studies \cite{mhamdi2024constraints} appear to have emphasized parameter constraints without a similarly detailed exploration of these cosmological diagnostics.}

In this paper, the $f(Q)$ modification of symmetric teleparallel gravity is studied. In Section 2, we start by going over the cosmological foundation of $f(Q)$ gravity. Our main contribution, which is given in Section 3, presents a particular exponential $f(Q)$ gravity model, explains how it is theoretically formulated, and considers its possible cosmological ramifications. Section 4 uses Markov Chain Monte Carlo (MCMC) techniques to constrain the model's parameters using a range of observational datasets in order to test it. After estimating the parameters, we examine the deceleration and equation of state (EoS) parameters, fractional energy densities, the $Om$ and Statefinder diagnostic tools, the energy conditions, the predicted age of the universe, and the model's behavior. Concluding remarks and a summary of the results are presented in Section 5.






\section{$f(Q)$ Gravity} \label{Sec:1}
In recent years, there has been significant interest in the non-metricity approach to gravity, particularly in its non-linear form known as $f(Q)$ gravity \cite{koussour2023thermodynamical,dimakis2025static}. Much of the current research explores its implications in cosmological settings and black hole physics. This focus is understandable, given that one of the primary motivations for investigating such non-linear models is their potential to account for various cosmic phenomena, such as dark energy, the inflationary field, and dark matter, without invoking additional components beyond the gravitational framework itself, as is required in standard General Relativity. In the following, we summarize the key findings related to the role of $f(Q)$ gravity \cite{yadav2024reconstructing,narawade2025insights} in cosmology. 

\subsection{Cosmology of $f(Q)$ gravity}\label{Sec:2}

The action for the $f(Q)$ theory of gravity, as formulated in \cite{Jimenez_2018_98}, is expressed as:

\begin{equation}\label{Eq:1}
S = \frac{1}{2} \int d^4x \sqrt{-g} f(Q) + S_m.
\end{equation}

Here, the natural units are chosen such that $\frac{8\pi G }{c^4} = 1$, and $S_m$ denotes the action corresponding to matter fields. When this action is varied with respect to the metric tensor, one obtains the following field equations:

\begin{equation}\label{Eq:2}
      \frac{2}{\sqrt{-g}}\nabla_\alpha \left( \sqrt{-g} f_Q P^\alpha_{~~\mu\nu} \right) - \frac{1}{2} f g_{\mu\nu} + f_Q \left( P_{\nu\alpha\beta} Q_{\mu}^{~~\alpha\beta} - 2 P_{\alpha\beta\nu} Q^{\alpha\beta}_{~~~\nu} \right) = T_{\mu\nu},
\end{equation}

where the derivative of the function $f(Q)$ with respect to the non-metricity scalar $Q$ is denoted by $f_Q = \frac{d f}{d Q}$. On the other hand, varying the action with respect to the affine connection gives the following field equations governing the dynamics of the connection:

\begin{equation}\label{Eq:3}
    \nabla_\mu \nabla_\nu \left( \sqrt{-g} f_Q P^{\mu\nu}_{~~~\alpha}\right) = 0,
\end{equation}

which serves as the equation of motion for the connection field. The tensor $ P^\alpha_{\mu\nu} $, which is conjugate to the non-metricity, is defined as:

\begin{equation}\label{Eq:4}
    P^{\alpha\mu\nu} = -\frac{1}{4} Q^{\alpha\mu\nu} + \frac{1}{2} Q^{(\mu\nu)\alpha} + \frac{1}{4} (Q^\alpha - \tilde{Q}^\alpha) g^{\mu\nu} - \frac{1}{4} g^{\alpha^{(\mu} Q^{\nu)}},
\end{equation}

where the symmetrization over indices is understood as $A_{(\mu\nu)} = \frac{1}{2} (A_{\mu\nu} + A_{\nu\mu})$. The two independent traces of the non-metricity tensor are introduced as:

\begin{equation}\label{Eq:5}
    Q_\alpha = g^{\sigma\lambda} Q_{\alpha\sigma\lambda}, \quad \tilde{Q}_\alpha = g^{\sigma\lambda} Q_{\sigma\alpha\lambda}.
\end{equation}

Finally, the non-metricity scalar $Q$ is constructed from a contraction of the non-metricity tensor and its conjugate:

\begin{equation}\label{Eq:6}
Q = -Q_{\alpha\mu\nu} P^{\alpha\mu\nu}.
\end{equation}
We adopt a spatially flat, homogeneous, and isotropic Friedmann–Lemaître–Robertson–Walker (FLRW) spacetime, described by the metric:

\begin{equation}\label{Eq:7}
ds^2 = -dt^2 + a^2(t) \left(dx^2 + dy^2 + dz^2\right),
\end{equation}
where $a(t)$ is the cosmological scale factor. This form of the metric assumes the coincident gauge, which effectively treats the metric as the only fundamental dynamical variable. Any deviation from this gauge would yield additional contributions to the gravitational field equations due to the affine connection, as discussed in \cite{Jimenez_2018_98,heisenberg_2023_review}. Under this specific choice of metric Eq.\eqref{Eq:7}, the non-metricity scalar $Q$ simplifies to:
\begin{equation}\label{Eq:8}
Q = 6H^2,
\end{equation}
with $H=\frac{\dot{a}}{a}$ being the Hubble parameter that characterizes the rate of cosmic expansion. The matter sector is modeled as a perfect fluid, whose energy-momentum tensor takes the form:
\begin{equation}\label{Eq:9}
T_{\mu\nu} = (\rho + p) u_\mu u_\nu + p g_{\mu\nu}.
\end{equation}
Here, $\rho$ and $p$ denote the fluid's energy density and pressure, respectively, and $\mu_\mu$ represents the four-velocity vector of the fluid. Substituting the FLRW metric Eq.\eqref{Eq:7} into the general field equations of $f(Q)$ gravity given by Eq.\eqref{Eq:2}, one obtains the cosmological equations:

\begin{eqnarray}\label{Eq:10}
6H^2~f_Q - \frac{1}{2}~f &=& \rho =\rho_m+\rho_r,  \\
-2\dot{H}(12H^2f_{QQ}+f_{Q})\ &=&\rho+p= \rho_m+p_m + \rho_r+p_r.\label{Eq:11}
\end{eqnarray}
where $\rho_m$, $p_m$ are the energy density and pressure of the matter, and $\rho_r$ and $p_r$ are the energy density and pressure of the radiation. Finally, the equations conclude with considering matter and radiation conservation equations:
\begin{equation}\label{Eq:010}
\dot{\rho}_m+3H(\rho_m+p_m)=0,
\end{equation}
\begin{equation}\label{Eq:0010}
\dot{\rho}_r+3H(\rho_r+p_r)=0.
\end{equation}
The universe is filled with dust and radiation, creating a dynamic environment, so that
\begin{equation}\label{Eq:15}
p_r=\frac{1}{3}\rho_r,\ \ \ \ p_m=0.
\end{equation}
The expressions Eq.\eqref{Eq:10} and Eq.\eqref{Eq:11} describe the evolution of the universe within the $f(Q)$ framework and form the foundation for further analysis. Now, introducing a redefinition of the function as $f(Q)=Q+\Psi(Q)$, the above equations \eqref{Eq:10} and \eqref{Eq:11} can be recast as:

\begin{equation}\label{Eq:12}
3H^2 = \rho_{total}=\rho + \frac{1}{2}\Psi - Q \Psi_Q,
\end{equation}

\begin{equation}\label{Eq:13}
-3H^2 - 2\dot{H} =p_{total}= p + Q \Psi_Q - \frac{1}{2} \Psi + 2\dot{H}(2Q \Psi_{QQ} + \Psi_Q),
\end{equation}
where $\rho$ and $p$ are defined in Eqs. \eqref{Eq:10} and \eqref{Eq:11}. These reformulated equations illustrate the deviation from standard general relativity due to the additional function $\Psi(Q)$, encapsulating modifications introduced by the $f(Q)$ theory.

\section{Exponential $f(Q)$ Gravity: Model Formulation and Cosmological Implications}
We now delve into a specific exponential model within the $f(Q)$ \cite{pradhan2024cosmological} gravity framework, systematically outlining its formulation and examining its key cosmological consequences. The choice of the function $f(Q)$ plays a pivotal role in enabling the accelerated expansion of the universe. A notable case is when $f(Q)=Q+2\Lambda$, which corresponds to General Relativity (GR) supplemented by a cosmological constant. Deviating from this by allowing a general form of 
$f(Q)$ introduces modifications to the Einstein-like field equations, particularly through additional terms that depend on derivatives of $f(Q)$. Among the various functional forms proposed in the literature, one particularly interesting case is the power-exponential model discussed by Anagnostopoulos et al. \cite{Anagnostopoulos_2021_822}, which demonstrates several noteworthy cosmological features. Another significant model is the power-law type examined in \cite{paliathanasis_2025}, where the resulting field equations can be recast into an integrable Hamiltonian form. Under specific parameter conditions, this approach offers an analytical expression for the dynamical dark energy equation of state (EoS) parameter. Early insights into the potential of $f(Q)$ gravity, a modification of general relativity arising from non-metricity proposed as an alternative to the conventional $\Lambda$CDM paradigm which is of the form $H(z) = H_0 \sqrt{\Omega_m(1+z)^3 + (1 - \Omega_m)}$ with a fixed matter density $\Omega_m = 0.3$ and $H_0$ being the present value of the Hubble parameter, were demonstrated in the study of  Anagnostopoulos, Basilakos, and Saridakis~\cite{Anagnostopoulos_2021_822}. Their study introduced a novel $f(Q)$ model and tested it against a diverse range of observational datasets, including cosmic chronometers (CC), baryon acoustic oscillations (BAO), and type Ia Supernovae (SNIa), demonstrating that $f(Q)$ gravity can effectively account for the observed cosmic expansion. Building on this foundation, Gadbail and Sahoo~\cite{gadbail2024modified} worked on a variety of $f(Q)$ models including, power-law, exponential, and logarithmic forms.  They~\cite{gadbail2024modified} examined the capability of such models to get hold of the evolution predicted by $\Lambda$CDM. The present work is directly inspired by the comprehensive observational analysis conducted by Mhamdi et al.~\cite{mhamdi2024constraints} that focused specifically on constraining power-law and exponential $f(Q)$ gravity models using observational data.

In the present work, we consider a class of well-defined exponential function of the form $\Psi(Q)$, given as
\begin{equation}\label{Eq013}
\Psi(Q) = \eta_1 Q_{0}\left(1-e^{-\eta_2\sqrt{\frac{Q}{Q_{0}}}}\right),
\end{equation}
where $\eta_1$ and $\eta_2$ are the model free parameters and $Q_0=6H_0^2$. Hence, our chosen $f(Q)$ gravity model is
\begin{equation}\label{Eq:013}
f(Q) = Q+\eta_1 Q_{0}\left(1-e^{-\eta_2\sqrt{\frac{Q}{Q_{0}}}}\right).
\end{equation}

\textcolor{black}{At this juncture, let us mention the motivation behind the choice of $f(Q)$ involving $\Psi(Q)$ in the above form. Three significant works \cite{mhamdi2024constraints,Anagnostopoulos_2021_822,gadbail2024modified} in this direction served as the initial motivation for such a choice. In order to explain physically, first of all, it contains an exponential term and for small values of $Q$, the exponential form can be expanded through Taylor series, which can obviously result in some moderate correction to GR. On the other hand, for large $Q$, the exponential term with negative sign will tend to zero, and in that case $f(Q) \to Q+\eta_1 Q_0$, which implies a small shift from standard GR. Moreover, it is a known fact that the exponential structure generates the scope of a smooth transition between the high curvature in the early universe and the low curvature in the late universe. This enables the study of cosmological dynamics with better phenomenological flexibility. We also know that this form of $f(Q)$ is analogous to well-established models of $f(R)$ and $f(T)$ gravity that have shown unification of early inflation and late-time acceleration.} \textcolor{black}{The present study considers a specific exponential form of symmetric teleparallel gravity given by \( f(Q) = Q + \eta_1 Q_0 \left(1 - e^{-\eta_2 \sqrt{Q/Q_0}} \right) \), designed to interpolate smoothly between early-time inflation and late-time acceleration. This form introduces nonlinear corrections to GR and is constrained using CC, Pantheon, and Pantheon$^+$ datasets via Markov Chain Monte Carlo (MCMC) analysis. However, Mhamdi et al. (2024) used a power-law form \( f(Q) = \alpha + \beta Q^n \) in their studies in ~\cite{mhamdi2024cosmological}.  This study used CC, Pantheon$^+$, and Redshift Space Distortion (RSD) data to constrain their model and investigated both linear (\( n=1 \)) and non-linear (\( n \neq 1 \)) regimes.  Further, an extended analysis by the same group in ~\cite{mhamdi2024observational} focused on growth index parameterizations (\( \gamma = \gamma_0 + \gamma_1 y(z) \)) within the same power-law framework of \( f(Q) \), treating the growth index as a free parameter. Thus, while both referenced works utilize a power-law \( f(Q) \) to evaluate growth dynamics and RSD data, the present work employs an exponential form tailored to capture dynamical transitions more smoothly, constrained with an updated multi-probe dataset. A different exponential-type structure was used by Vasquez and Oliveros \citep{vasquez2025}, who proposed \( f(Q) = Q + 2\Lambda \exp\left[ -(b\Lambda/Q)^n \right] \), which is a perturbative expansion around the \( \Lambda \)CDM model. The method adopted therein allows cosmological parameters to be treated analytically and explores how the model behaves with various parameter selections, including the impact on matter perturbations. Although the goal of both models reported here and \citep{vasquez2025} is to generalize GR with exponential corrections, the current model uses a multi-probe numerical MCMC framework and introduces corrections through a square-root dependence in the exponent, while the \( f(Q) \) form in \citep{vasquez2025} is designed to approximate \( \Lambda \)CDM in a controlled perturbative manner, enabling analytical solutions under small-deviation assumptions. Let us further note that \(f(Q) = \frac{Q}{8\pi G} - \alpha \ln\left(\frac{Q}{Q_0}\right)\) is the logarithmic model suggested by \cite{Najera_2023_524}. Through a logarithmic correction, this form introduced a departure from the symmetric teleparallel equivalent of General Relativity (STEGR). Crucially, this form provides a possible solution to the cosmological constant problem by avoiding the explicit inclusion of a constant term. Although the cosmic acceleration is geometrically induced by both the current model and the model of \cite{Najera_2023_524}, their theoretical justifications and phenomenological implications are different. The exponential model provides more flexibility in capturing dynamical transitions across cosmic epochs, whereas the logarithmic model stresses minimal coupling and analytical tractability, especially in perturbative regimes. Based on all this, we can say that such a choice of $\Psi(Q)$ results in the form of an $f(Q)$ that is suitable for exploring the late-time acceleration as well as early inflationary cosmology.}

The first and second derivatives of $f(Q)$ with respect to the non-metricity scalar $Q$ are:
\begin{equation}\label{Eq:0013}
f_Q = 1+\frac{\eta_1\eta_2}{2}\sqrt{\frac{Q_0}{Q}}e^{-\eta_2\sqrt{\frac{Q}{Q_0}}},
\end{equation}
\begin{equation}\label{Eq:00013}
f_{QQ} = -\frac{\eta_1\eta_2}{4Q}e^{-\eta_2\sqrt{\frac{Q}{Q_0}}}\left(\sqrt{\frac{Q_0}{Q}}+\eta_2\right).
\end{equation}

Along with utilizing numerical methods to frame the cosmological model, a specific set of dimensionless and model-independent variables is examined as

\begin{equation}\label{Eq:14}
\mu=\frac{\rho_m}{3H^2},\ \ \ \nu=\frac{\rho_r}{3H^2}\ \ \ \zeta=\frac{H}{H_0}.
\end{equation}

Now, using Eq.\eqref{Eq:14}, we have the density parameters for matter, radiation, and dark energy, respectively, from Eq.\eqref{Eq:12} as
\begin{equation}\label{Eq:16}
\begin{array}{cc}
\Omega_m=\mu, ~\Omega_r=\nu=\frac{e^{-\eta_2  \zeta} \left(\eta_1 +\eta_1  \eta_2  \zeta-e^{\eta_2  \zeta} \left(\eta_1 +\zeta^2 (-1+\mu)\right)\right)}{\zeta^2}, \\
\\
\Omega_{f}=\frac{e^{-\eta_2  \zeta} \eta_1  \left(-1+e^{\eta_2  \zeta}-\eta_2  \zeta\right)}{\zeta^2}.
\end{array}
\end{equation}
From the relation $H(z) = -\frac{1}{1+z} \frac{dz}{dt}$ and by taking derivatives of $\mu$ and $\zeta$ with respect to the redshift $z$,  we can obtain two coupled differential equations using Eqs.~\eqref{Eq:13},~\eqref{Eq:14} and \eqref{Eq:16} as
    \begin{equation}\label{Eq:20}
    \frac{d\mu}{dz}=\frac{-\eta_1  (8+\eta_2  \zeta (8+3 \eta_2  \zeta)) \mu +2 e^{\eta_2  \zeta} \left(4 \eta_1 +\zeta^2 (-1+\mu)\right) \mu}{(1+z) \left(2
e^{\eta_2  \zeta}-\eta_1  \eta_2 ^2\right) \zeta^2},
\end{equation}
\begin{equation}\label{Eq:21}
    \frac{d\zeta}{dz}=\frac{4 (\eta_1 +\eta_1  \eta_2  \zeta)-e^{\eta_2  \zeta} \left(4 \eta_1 +\zeta^2 (-4+\mu)\right)}{(1+z) \left(2 e^{\eta_2 \zeta}-\eta_1
 \eta_2 ^2\right) \zeta}.
\end{equation}
The above differential equations encompass the redshift evolution of $\mu(z)$ and $\zeta(z)$ under the exponential $f(Q)$ gravity scenario. The structure of Eqs.~(\ref{Eq:20}) and (\ref{Eq:21}) is inherently nonlinear due to the presence of polynomial and exponential terms in $\zeta$ that are rooted from the underlying functional form of the $f(Q)$ Lagrangian. The coupling between $\mu$ and $\zeta$ implies a dynamical interplay that governs the background cosmological evolution. These equations are particularly suitable for numerical integration, allowing for direct comparison with observational datasets such as CC \cite{lu2023observational}, Pantheon \cite{myrzakulov2024observational}, and Pantheon$^+$ \cite{brout2022pantheon}. Their behavior across redshift is supposed to provide a critical insight into the viability and consistency of the adopted $f(Q)$ model in describing the late-time accelerated expansion of the universe.

This system of coupled differential equations (\ref{Eq:20}) and (\ref{Eq:21}) is numerically solved using initial conditions $\mu_0=\Omega_{m_0}$ and $\zeta_0 = 1$. In the context of the adopted $f(Q)$ gravity model, this framework introduces four independent free parameters, such as the present-day Hubble constant $H_0$, the current matter density parameter $\Omega_{m_0}$, and the model parameters $\eta_1$ and $\eta_2$. In the following section, we will utilize cosmological observational datasets to constrain these parameters.

\section{Constraining the Model Parameters through Observational Data}

In this section, we use recent cosmological datasets to constrain the parameters of the adopted $f(Q)$ gravity model in order to evaluate its feasibility. CC \cite{moresco2012improved,moresco2015raising}, Type Ia Supernovae from the Pantheon \cite{scolnic2018complete,Najera_2021_34} compilation, and its extended version, Pantheon$^+$ \cite{brout2022pantheonplus,scolnic2022pantheonplus}, are among the observational data used. These datasets offer strong constraints on the universe's expansion history, making it possible to estimate the model parameters consistently. The details are being elaborated in the subsequent subsections. 

\subsection{Cosmological Parameter Estimation Using MCMC Techniques}
To constrain the cosmological parameters $\boldsymbol{\Theta} = (H_0, \Omega_{m_0}, \eta_1, \eta_2)$ 
we adopt a Bayesian statistical inference framework \cite{foreman2013emcee,karamanis2021zeus} supported by Markov Chain Monte Carlo (MCMC) \cite{Lewis2002, foreman2013emcee} sampling. This method allows for efficient exploration of the parameter space and robust estimation of the posterior probability distributions of cosmological parameters. According to Bayes’ theorem, the posterior distribution is given by:
\begin{equation}\label{O1}
    P(\boldsymbol{\theta} \mid D, I) = \frac{P(\boldsymbol{\theta} \mid I) \cdot P(D \mid \boldsymbol{\theta}, I)}{P(D \mid I)},
\end{equation}
where $P(\boldsymbol{\theta} \mid I)$ represents prior knowledge of the parameters, $P(D \mid \boldsymbol{\theta}, I)$ is the likelihood of the observed data $D$ given the model defined by $\boldsymbol{\theta}$, and $P(D \mid I)$ is the Bayesian evidence (a normalization constant). The likelihood function is expressed in terms of the chi-squared statistic:
\begin{equation}\label{O2}
\mathcal{L}(\boldsymbol{\theta}) = \exp\left(-\frac{\chi^2(\boldsymbol{\theta})}{2}\right).
\end{equation}
We apply this methodology to three major observational datasets: the cosmic chronometers (CC) data, the Pantheon Type Ia Supernovae sample, and the extended Pantheon$^+$ compilation.

\subsubsection*{\textbf{Cosmic Chronometers $(CC)$ Dataset:}}
The cosmic chronometers \cite{Moresco_2022_25} technique provides direct, model-independent measurements of the Hubble parameter $H(z)$ by using the relative ages of massive, passively evolving galaxies. This approach relies on the relation:
\begin{equation}\label{O3}
    H(z) = -\frac{1}{1+z} \frac{dz}{dt},
\end{equation}
where $dz/dt$ is derived from differential age evolution across redshift, measured from early-type galaxies that formed at high redshifts and have since evolved passively. Because this method does not depend on a specific cosmological model, it serves as a powerful tool for independently constraining the expansion history of the universe.

In this analysis, we use a compilation of 32 CC data points \cite{Borghi_2022_928}, covering the redshift range $0.07 \leq z \leq 1.965$, obtained from various spectroscopic surveys. These data are associated with standard errors and offer sparse but crucial coverage of $H(z)$ at intermediate redshifts. The corresponding chi-squared function used to constrain the model parameters is:
\begin{equation}\label{O4}
\chi^2_{\text{CC}}(\boldsymbol{\theta}) = \sum_{i=1}^{32} \frac{\left[H_{\text{th}}(z_i; \boldsymbol{\theta}) - H_{\text{obs}}(z_i)\right]^2}{\sigma_H^2(z_i)},
\end{equation}
where $H_\text{th}$ and $H_\text{obs}$ are the theoretical and observational values of the Hubble parameter for different values of redshifts, respectively, and $\sigma_H$ is the standard error.

\subsubsection*{\textbf{Pantheon Supernovae Dataset}}

The Pantheon dataset is one of the largest compilations of Type Ia Supernovae (SNe Ia) \cite{shekh2023new,shukla2024cosmic}, comprising 1048 SNe in the redshift range $0.01 < z < 2.3$. This dataset combines several high-quality subsamples including those from the Sloan Digital Sky Survey (SDSS), the Supernova Legacy Survey (SNLS), Pan-STARRS1 Medium Deep Survey \cite{scolnic2018complete}, and low-redshift surveys. SNe Ia are considered “standardizable candles” due to their relatively uniform intrinsic brightness and their well-calibrated empirical relations (e.g., light curve shape and colour corrections). The key observable from SNe Ia is the distance modulus, defined as:
\begin{equation}\label{O5}
\mu_{\text{th}}(z; \boldsymbol{\theta}) = 5 \log_{10} \left[d_L(z; \boldsymbol{\theta})\right] + \mu_0,
\end{equation}
where $\mu_0$ is a nuisance parameter combining the absolute magnitude and the Hubble constant. The luminosity distance $d_L(z)$ is computed as:
\begin{equation}\label{O6}
d_L(z) = (1 + z) \int_0^z \frac{dz'}{E(z')},
\end{equation}
where $E(z) = \frac{H(z)}{H_0}$ is the normalized Hubble parameter and variable $z'$ denotes the integration parameter ranging from $0$ to redshift $z$. The chi-squared function for the Pantheon sample is then:
\begin{equation}\label{O7}
\chi^2_{\text{SN}}(\boldsymbol{\theta}) = \sum_{i=1}^{1048} \frac{\left[\mu_{\text{th}}(z_i; \boldsymbol{\theta}) - \mu_{\text{obs}}(z_i)\right]^2}{\sigma^2_\mu(z_i)},
\end{equation}
where $\mu_{\text{th}}$ and $\mu_{\text{obs}}$ denote the theoretical and observed distance moduli respectively and $\sigma_\mu$ denotes their uncertainties.

\subsubsection*{\textbf{Pantheon$^+$ Supernovae Dataset} }

The Pantheon$^+$ sample \cite{brout2022pantheon,scolnic2022pantheonplus,brout2022pantheonplus} is a comprehensive and updated successor to the original Pantheon compilation. It includes 1701 light curves of 1550 unique SNe Ia, offering improved systematics, calibration, and redshift coverage. The dataset spans a redshift interval of $0.00122 \leq z \leq 2.2613$ and combines data from ground-based and space-based observations, such as Hubble Space Telescope (HST) programs. Pantheon$^+$ provides refined treatment of systematic errors, updated calibration of zero-points, and new host-galaxy corrections. These enhancements improve the reliability of cosmological constraints derived from SNe Ia and increase sensitivity to late-time cosmic acceleration. As with the Pantheon dataset, the chi-squared function for Pantheon$^+$ is:
\begin{equation}\label{O8}
\chi^2_{\text{SN}}(\boldsymbol{\theta}) = \sum_{i=1}^{1701} \frac{\left[\mu_{\text{th}}(z_i; \boldsymbol{\theta}) - \mu_{\text{obs}}(z_i)\right]^2}{\sigma^2_\mu(z_i)}.
\end{equation}

\subsubsection*{\textbf{Combined Chi-Squared and MCMC Sampling}}
In cases where both CC and SNe Ia data are used, we define the total chi-squared as the sum:
\begin{equation}\label{O9}
\chi^2_{\text{total}}(\boldsymbol{\theta}) = \chi^2_{\text{CC}}(\boldsymbol{\theta}) + \chi^2_{\text{SN}}(\boldsymbol{\theta}).
\end{equation}
After discarding an initial burn-in phase, we extract the marginalized posteriors and confidence intervals for $H_0$, $\Omega_{m_0}$, $\eta_1$, and $\eta_2$. 
The resulting distributions provide both the best-fit values and credible intervals, allowing a complete and statistically consistent interpretation of the underlying cosmological model.

\subsubsection*{\textbf{Corner Plots of Marginalized Posterior distributions for $H_0$, $\Omega_0$, $\eta_1$ and $\eta_2$}}

\begin{figure}[htbp]
\centering
\includegraphics[width=1.15\textwidth]{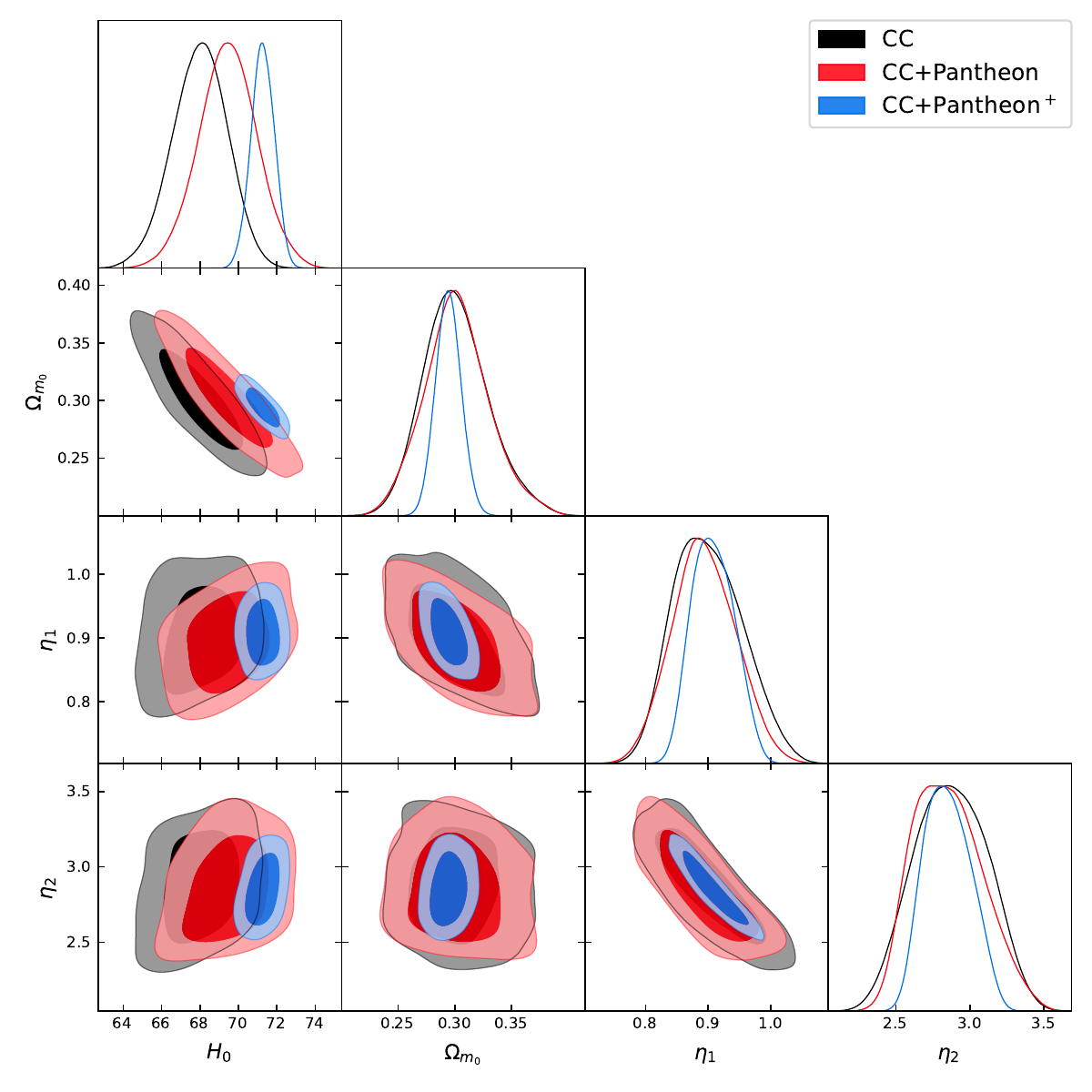}
\caption{Corner plot showing the marginalized posterior distributions with 68\% and 98\% confidence contours for the model parameters $H_0$, $\Omega_{m_0}$, $\eta_1$ and $\eta_2$. The contours are obtained for CC (black), CC+Pantheon (red), CC+Pantheon$^+$(blue). \label{fig1}}
\end{figure}

In Figure~\ref{fig1}, we have displayed the corner plot that represents the outcomes of MCMC analysis for the parameter space $\Theta=(H_0,\Omega_{m_0},\eta_1,\eta_2)$, where $H_0$, $\Omega_{m_0}$, $\eta_1$, and $\eta_2$ represent the current value of $H$, the current matter density parameter and free model parameters, respectively. In this analysis, we are taking into account three different observational data sets that include CC, CC+Pantheon, and CC+Pantheon$^+$ represented by black contours, red contours, and blue contours, respectively. The marginalized posterior means and the uncertainties ($1\sigma$ indicated by dark regions and $2\sigma$ indicated by lighter regions) for the observational data sets are displayed in the Figure~\ref{fig1}. If we observed the contour plot reveals that the posterior distributions have been narrowed significantly due additional data points into the $Pantheon$ and its a $Pantheon^+$ dataset. Moreover, it is noticeable that this narrowing is most prominently visible for $H_0$ and $\Omega_{m_0}$. Furthermore, in general, there is good overlap among the datasets, and this indicates the consistency of the model with observation. The blue contours also make it clear that there are tighter constraints corresponding to it. Although $\eta_1$ and $\eta_2$ appear to be stable across the datasets, a relatively better stability is apparent for $\eta_2$ 
where we observe nearly overlapping distributions. For $\eta_1$, although we observe good overlap of the distributions across the datasets, a mild sensitivity is apparent through slightly shifted contours due to tension. Combining all these observations we can interpret that incorporation of additional datasets significantly improves constraints on the cosmological parameters and helps in reducing the degeneracies. In general, this corner plot appears to exhibit robustness of the theoretical model under consideration.

\begin{figure}[htbp]
\centering
\includegraphics[width=1.15\textwidth]{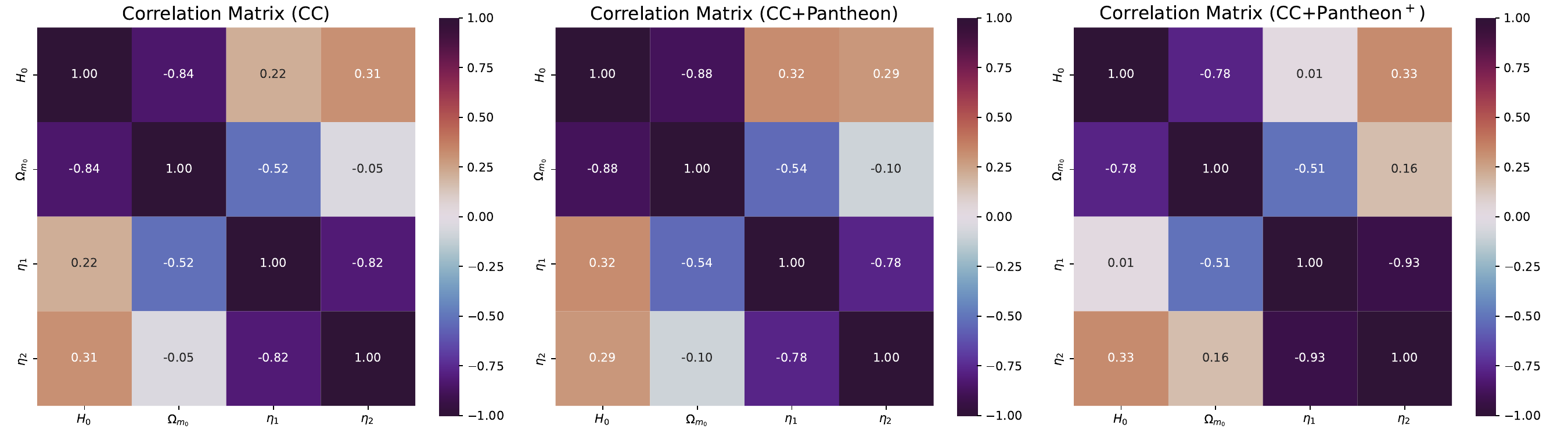}
\caption{Correlation matrices for the model parameters \( H_0 \), \( \Omega_{m_0} \), \( \eta_1 \), and \( \eta_2 \), obtained from the CC, CC+Pantheon, and CC+Pantheon$^+$ dataset combinations. \label{fig01}}
\end{figure}

The correlation matrices presented in Figure~\ref{fig01} illustrate the mutual dependencies among the cosmological parameters \( H_0 \) (Hubble constant), \( \Omega_{m_0} \) (present-day matter density parameter), and the model-specific parameters \( \eta_1 \) and \( \eta_2 \)  for all dataset combinations. In all the cases, a strong negative correlation is evident between \( H_0 \) and \( \Omega_{m0} \), reflecting the well-established degeneracy between the expansion rate and matter content in low-redshift cosmological observations~\cite{aghanim2020planck}. The parameters \( \eta_1 \) and \( \eta_2 \) consistently exhibit a strong negative correlation, indicative of a compensatory mechanism intrinsic to the extended model framework. It may be noted with importance that the inclusion of the Pantheon$^+$ dataset significantly weakens the correlation between \( H_0 \) and \( \eta_1 \). This suggests that the improved Supernovae constraints provided by Pantheon$^+$  effectively disentangle the expansion history from model-specific modifications. At the same time, the correlation between \( \eta_1 \) and \( \eta_2 \) comes out to be even more prominent, indicating a heightened sensitivity of the model to their joint variation in the presence of higher-precision luminosity distance data.

\begin{figure}[htbp]
\centering
\includegraphics[width=1.1\textwidth]{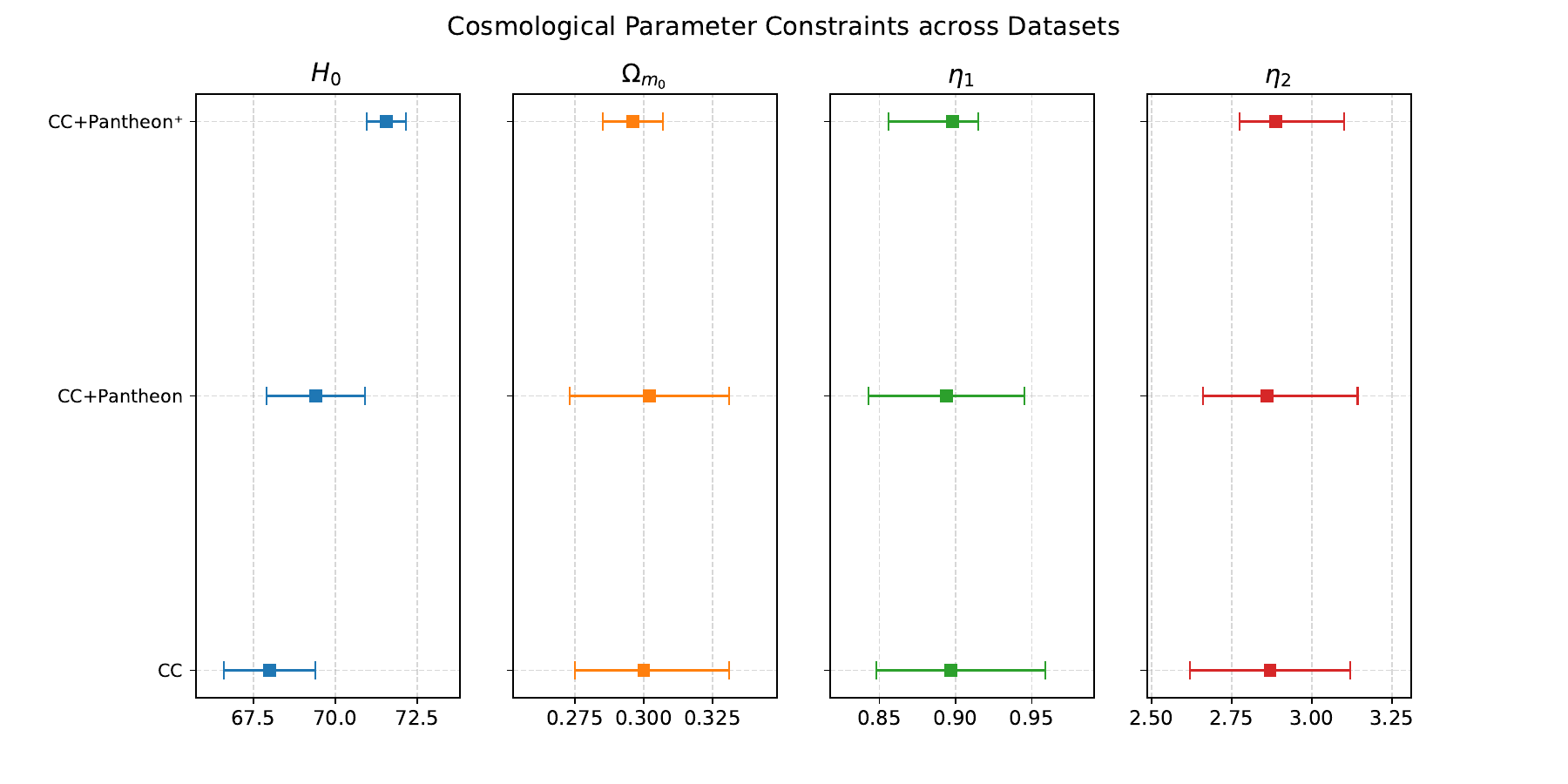}
\caption{Whisker plot showing the constraints on cosmological parameters ($H_0$, $\Omega_{m0}$, $\eta_1$, and $\eta_2$) from different datasets (CC, CC+Pantheon, and CC+Pantheon$^+$).}
\label{fig02}
\end{figure}

The parameter constraints are exhibited in Table~\ref{table:1}. It is apparent from the Table~\ref{table:1} that improved precision is available with the inclusion of Pantheon and Pantheon$^+$ datasets, especially for $H_0$ and $\Omega_{m_0}$.  It may be noted that, the $CC + \text{Pantheon}^+$ combination results in more efficient constraining, indicating enhanced sensitivity and model stability from multi-probe data fusion. The current values of key cosmological parameters obtained from the exponential $f(Q)$ gravity model using the CC, CC+Pantheon, and CC+Pantheon$^+$ datasets are shown in Table~\ref{table:2}. The values appear to be consistent with late-time acceleration and are in alignment with observational estimates. The whisker plot for the adopted $f(Q)$ gravity model is shown in Figure~\ref{fig02}. This plot provides a visual representation of the distributions of the model parameters constrained for different datasets. The whisker plot displaying the constraints on the cosmological parameters $H_0$, $\Omega_{m_0}$, $\eta_1$, and $\eta_2$ obtained from the CC, CC+Pantheon, and CC+Pantheon$^+$ datasets is shown in Figure~\ref{fig02}. The CC dataset for $H_0$ produces a value of about $68.5 \pm 0.5$ km s$^{-1}$ Mpc$^{-1}$, whereas the addition of Pantheon data (CC+Pantheon and CC+Pantheon$^1$) causes the central value to shift to about $70.0 \pm 0.5$ km s$^{-1}$ Mpc$^{-1}$, indicating a tighter constraint and a higher preferred value for the Hubble constant. All three datasets consistently constrain the matter density parameter to roughly $0.30 \pm 0.01$ in the case of $\Omega_{m_0}$. These demonstrate agreement across the various data combinations. Similarly, the parameters $\eta_1$ and $\eta_2$ show strong constraints, with $\eta_1$ tightly clustered around $0.90 \pm 0.01$ and $\eta_2$ around $2.85 \pm 0.05$. Interestingly, adding the Pantheon Supernovae data generally reduces or maintains the uncertainties for all parameters, demonstrating the enhanced precision provided by the combined datasets, especially for $H_0$. Although there is a slight tension in the $H_0$ values across datasets, which is a common feature in contemporary cosmological analyses, the consistency across datasets for $\Omega_{m_0}$, $\eta_1$, and $\eta_2$ indicates that the selected $f(Q)$ model is well-constrained by the current observational data.

\begin{table}[hbt]
\renewcommand\arraystretch{1.5}
\centering 
\begin{tabular}{||c|c|c|c|c||} 
\hline\hline 
~~~Dataset~~~&~~~~~~~ $H_{0}$ ~~~~~~~& ~~~~~~~$\Omega_{m_0}$~~~~~~~ & ~~~~~~~~~~$\eta_1$~~~~~~~&~~~~~~~$\eta_2$~~~~~~\\ [0.5ex]  
\hline\hline
CC & $68\pm 1.4$ &   $0.300^{+0.025}_{-0.031}$ & $0.897^{+0.049}_{-0.062}$ & $2.87 \pm 0.25$ \\[0.5ex]
\hline
CC+Pantheon & $69.4\pm 1.5$ &  $0.302\pm0.029$ & $0.894\pm0.051$ & $2.861^{+0.201}_{-0.282}$\\[0.5ex]
\hline
CC + Pantheon$^+$ & $71.56\pm 0.60$ &  $0.296 \pm 0.011$ & $0.898^{+0.042}_{-0.017}$ & $2.887^{+0.112}_{-0.215}$\\[0.5ex]
\hline \hline 
\end{tabular}
\caption{Constrained parameter values for the \textcolor{black}{fitted} $f(Q)$ gravity model are obtained using three observational datasets: CC, CC + Pantheon, and CC + Pantheon$^+$.}
\label{table:1} 
\end{table}

\begin{table}[htb]
\renewcommand{\arraystretch}{1.5}
\centering
\begin{tabular}{||c|c|c|c||}
\hline\hline
Cosmological Parameter & CC & CC + Pantheon & CC + Pantheon$^+$ \\
\hline\hline
$q_0$  & $-0.431$ & $-0.423$ & $-0.437$ \\
\hline
$z_{\mathrm{tr}}$ & $0.691$ & $0.671$ & $0.697$ \\
\hline
$\omega_0$  & $-0.620$ & $-0.615$ & $-0.624$ \\
\hline
$s_0$  & $0.088$ & $0.089$ & $0.085$ \\
\hline
$r_0$  & $0.753$ & $0.752$ & $0.760$ \\
\hline
Age of the universe [Gyr] & $13.888$ & $13.573$ & $13.357$ \\
\hline\hline
\end{tabular}
\caption{Best-fit present-day values of selected cosmological parameters for the \textcolor{black}{observationally-constrained} $f(Q)$ gravity model, obtained using three observational datasets: CC, CC + Pantheon, and CC + Pantheon$^+$.}
\label{table:2}
\end{table}

\subsubsection*{\textbf{Effect of Combined Data on $H(z)$ and $\Lambda$CDM Comparison}}

\begin{figure}[htbp]
\centering
\includegraphics[width=1.1\textwidth]{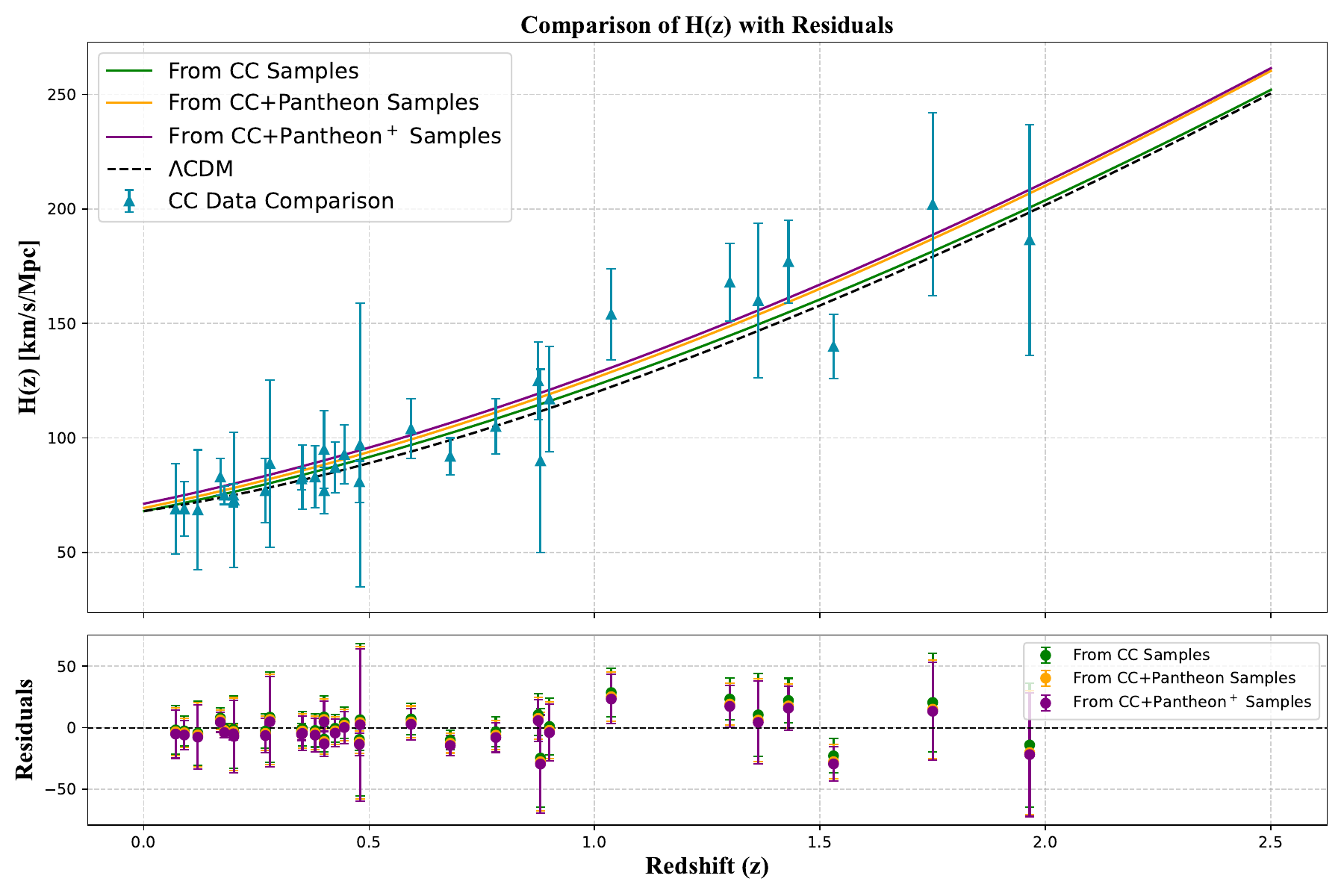}
\caption{Comparison of the Hubble parameter \( H(z) \) from different observational datasets with theoretical predictions. The upper panel shows \( H(z) \) measurements from CC, CC combined with Pantheon Supernovae, and CC + Pantheon$^+$, overlaid with the \(\Lambda\)CDM model. The lower panel displays the residuals for each dataset, highlighting deviations from the observation across redshift. Error bars represent standard uncertainties.}
\label{fig2}
\end{figure}
Figure~\ref{fig2} (Upper Panel) shows a comparison of the expansion function $H(z)$ with error bar plots for 32 point of CC data set (teal blue) with standard error in the observed value against redshift. Purpose of this plot is to emphasize on the effect of employing the three different combinations of observational data on the model predictions. The three solid lines refer to reconstructed $H(z)$ based on parameter sets derived from: (i) CC samples alone (green), (ii) CC + Pantheon Type Ia Supernovae sample (orange), and (iii) CC + Pantheon$^+$ samples (magenta). The black dashed line is the standard $\Lambda$CDM model, which is a reference for a flat universe with a cosmological constant. It is observed that the magenta line (CC+Pantheon$^+$) yields a somewhat higher $H(z)$ at higher redshifts ($z > 1.5$). We get an indication of a mild deviation from $\Lambda$CDM in this line possibly implying  a reflection of addition of the Pantheon$^+$ data points. Theoretical predictions for all three parameter sets are compatible with these measurements within errors over the entire redshift interval considered ($0 < z < 2.5$). The plot from CC data set (green) follow $\Lambda$CDM which totally derive from CC observational data points. However, the  lines, where complementary data like Pantheon and Pantheon$^+$, are showing mild deviation from $\Lambda$CDM with observational points. This behavior highlights the impact of using complementary data like Pantheon and Pantheon$^+$. This comparison reflects the necessity of multi-probe cosmological measurements. The convergence of results of various parameter sets to the $\Lambda$CDM prediction at lower redshifts, and the slight deviations at higher redshifts, shows the necessity for more accurate high-$z$ measurements.

In Figure~\ref{fig2} (Lower Panel), we computed the residuals of the Hubble parameter $H(z)$ using the 32-point CC dataset to quantitatively evaluate how well our cosmological model fits the observational data. For each redshift point $z_i$, the residual is defined as $\Delta H(z_i) = H_{\text{obs}}(z_i) - H_{\text{model}}(z_i)$, where $H_{\text{obs}}(z_i)$ is the observed Hubble parameter from the CC dataset and $H_{\text{model}}(z_i)$ is the model-predicted value computed from three sets of best-fit parameters: (i) using CC data alone, (ii) using combined CC and Pantheon data, and (iii) using CC with Pantheon$^+$ Supernovae. By plotting these residuals as a function of redshift, we assess the deviation of each model from the data. A good model fit should yield residuals scattered randomly around zero without showing any systematic trend. In our results, the CC model yields the most balanced and least structured residual distribution, indicating a superior agreement with the observational data compared to the other two configurations.

\textcolor{black}{Figure~\ref{fig2}'s lower panel (residuals) indicates observed departures from the $\Lambda$CDM model at high redshifts. In particular, the CC Samples and CC Data Comparison for $z \gtrsim 1.5$ show consistently negative residuals, suggesting that the measured Hubble parameter values $H(z)$ are consistently lower than the standard $\Lambda$CDM prediction. Physical interpretations of such high-redshift deviations generally include possible systematic effects or new physics beyond $\Lambda$CDM. Regarding the former, an evolving dark energy equation of state, $w(z)$, deviating from the constant $w = -1$ (e.g., $w > -1$ at high redshifts, characteristic of quintessence-like behavior) could lead to a slower expansion rate in the past compared to $\Lambda$CDM, thereby explaining the observed lower $H(z)$ values. As an alternative, models in which dark energy interacts with neutrinos or dark matter might alter the evolution of their energy density, which could impact the early expansion of the universe and possibly slow it down more effectively than $\Lambda$CDM. Poulin et al. \cite{poulin2019early} proposed a scalar field active at \( z \sim 3000 \), inducing a deviation from the \( \Lambda \)CDM model during recombination. Di Valentino et al. \cite{di2021realm} reviewed a wide range of high-\( z \) solutions including modified gravity, early dark energy (EDE), and interacting dark energy. Reference \cite{rameez2021there} discussed how early universe deviations (high-\( z \)) may be favored depending on analysis priors and data combinations. The observed discrepancies can also be explained by modified gravity theories that change effective energy densities or gravitational strengths at high redshifts \cite{odintsov2025modified, chaudhary2023parametrization}. In addition, a non-zero spatial curvature ($\Omega_k \neq 0$), especially a positively curved universe ($\Omega_k > 0$), may lead to a lower $H(z)$ at high redshifts than a flat $\Lambda$CDM model. In this context let us note the work of \cite{handley2021curvature}, who demonstrated through Bayesian evidence that \textit{Planck} data alone favor a positively curved universe, and shows that a closed model can reduce \( H(z) \) at high redshift. Another notable study in this context is done by \cite{di2020planck}, that report that \textit{Planck} CMB data prefer a closed universe at more than 99\% confidence level, which has implications for lowering \( H(z) \) at high redshifts. Regarding systematic effects and data limitations, the derivation of $H(z)$ at high redshifts, particularly from cosmic chronometers, depends on precise spectroscopic measurements and stellar population modeling; biases may be introduced by unreported systematic errors or calibration problems in these processes. The deviations that we observed between ``CC Samples" and ``CC+Pantheon Samples" at high redshifts indicate that the inclusion of Pantheon data or methodological differences might result in some of these systematics. Finally, sample variance or selection biases in the specific high-redshift samples used might also be responsible for the observed deviations. Some pertinent studies in this context are \cite{solanki2024bulk,lohakare2025late}. }

\subsubsection*{\textcolor{black}{\textbf{Model Comparison and Statistical Viability through AIC and BIC}}}
\textcolor{black}{In this subsection, we present an assessment of the viability of competing cosmological models. For that purpose we employ the Akaike Information Criterion (AIC) and Bayesian Information Criterion (BIC), which penalize model complexity to avoid overfitting. In both the kinematic analysis using SNe Ia and $H(z)$ data \citep{benndorf2022determination}, and the calibration of GRB Amati relations with Gaussian Process-reconstructed data \citep{han2024detection}, these criteria consistently favor simpler models, that highlighted the effectiveness of AIC and BIC in guiding cosmological model selection toward parsimony and statistical robustness. To compare how well the proposed exponential \( f(Q) \) gravity model performs statistically, we use the AIC and BIC. These criteria help identify models that fit the data well without being too complex by penalizing those with too many parameters \cite{cavanaugh2019akaike, mangan2017model}. Reference \cite{rezaei2021comparison} elaborately demonstrated how in cosmology, the AIC and the BIC are widely utilized for model selection, with the aim of balancing goodness of fit and model complexity. While AIC tends to favor models with better predictive accuracy, BIC applies a stronger penalty for additional parameters and thus often prefers simpler models such as $\Lambda$CDM over dynamical dark energy scenarios. Ref. \cite{arevalo2017aic} investigated linear and nonlinear cosmological interactions dependent on dark matter and dark energy densities within general relativity, utilizing AIC and BIC with SnIa, H(z), BAO, and CMB data to compare interacting models and identify those alleviating the coincidence problem.
The AIC and BIC are defined respectively as \cite{rezaei2021comparison}:
\begin{align}
\text{AIC} &= \chi^2_{\text{min}} + 2k, \\
\text{BIC} &= \chi^2_{\text{min}} + k \ln N,
\end{align}
where $k$ is the number of free parameters in the model and $N$ is the number of data points in the corresponding observational dataset.}

\begin{figure}[htbp]
\centering
\includegraphics[width=1.0\textwidth]{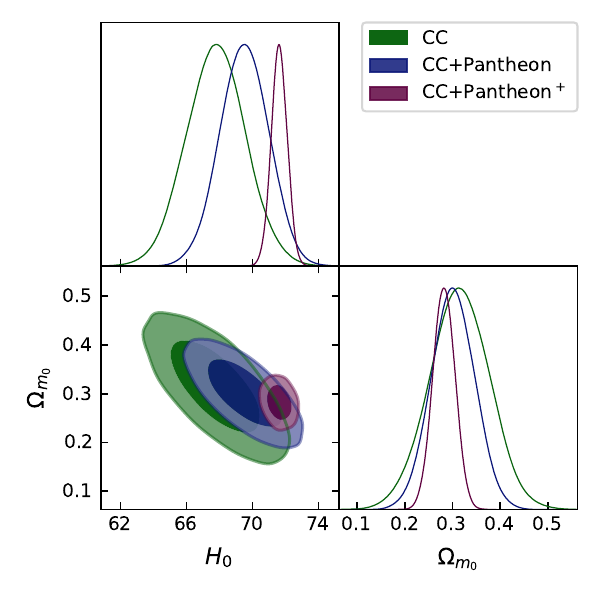}
\caption{\textcolor{black}{Corner plot showing the marginalized posterior distributions with 68\% and 98\% confidence contours for the parameters $H_0$ and $\Omega_{m_0}$ of the $\Lambda$CDM model. The contours are obtained for CC (green), CC+Pantheon (blue), and CC+Pantheon$^+$ (maroon).} \label{f1}}
\end{figure}

\textcolor{black}{In the first phase of this analysis, we have presented a corner plot in Figure~\ref{f1} for the $\Lambda$CDM model depicting marginalized posterior distributions for 68 \% and 98 \% confidence levels with the parameters $H_0$ and $\Omega_{m_0}$ utilizing CC (green), CC+Pantheon (blue) and CC+Pantheon$^+$ (maroon). For the CC data, $H_0$ and $ \Omega_{m_0}$ are aligned with Planck 2018 results. The other two datasets significantly narrowed the posterior distributions and the confidence contours. As a whole, the figure demonstrates the robustness of the $\Lambda$CDM model across the observational datasets and also indicates the importance of combining CC and Supernovae data for improved cosmological parameter estimation. In the next phase, we have carried out AIC and BIC analysis to assess the constrained $f(Q)$ model under consideration with the $\Lambda$CDM model in terms of AIC  and BIC.}

\textcolor{black}{To assess the statistical performance of the exponential $f(Q)$ gravity model relative to the standard $\Lambda$CDM cosmology, we employ three key information criteria: the minimum chi-squared ($\chi^2_{\rm min}$), the AIC, and the BIC. These metrics are computed for each model using three combinations of observational datasets as already mentioned i.e.: CC, CC combined with Pantheon, and CC combined with Pantheon$^+$. Table~\ref{T1} summarizes the best-fit cosmological parameters and statistical indicators for the $\Lambda$CDM model. The Hubble constant $H_0$ and matter density parameter $\Omega_{m_0}$ are constrained with increasing precision as more data are incorporated. Specifically, the inclusion of Pantheon$^+$ data yields $H_0 = 71.62^{+0.47}_{-0.49}$ km s$^{-1}$ Mpc$^{-1}$ and $\Omega_{m0} = 0.282^{+0.021}_{-0.024}$, with a corresponding $\chi^2_{\rm min} = 1647.89$. The AIC and BIC values for this dataset combination are 1651.89 and 1662.81, respectively.}

\textcolor{black}{If we look into Table~\ref{table:01}, we see that it presents the comparative model selection statistics for the exponential $f(Q)$ gravity model. The differences in AIC ($\Delta$AIC) and BIC ($\Delta$BIC) are computed relative to the $\Lambda$CDM. For all dataset combinations, the exponential $f(Q)$ model exhibits higher AIC and BIC values, indicating a statistical preference for the $\Lambda$CDM model under these criteria. However, the $\Delta$AIC values appear to be within the range $5 < \Delta \mathrm{AIC} < 15$. Hence, there is an indication that the exponential $f(Q)$ model under study is still moderately supported by the data. The statistical analysis reveals that while the exponential $f(Q)$ gravity model provides a viable fit to the data, the $\Lambda$CDM model remains statistically favored, particularly when evaluated using BIC, which imposes a stronger penalty for model complexity. It may be noted that $\Delta$AIC and $\Delta$BIC are little higher than the respective values as obtained by \cite{mhamdi2024cosmological}. According to the model selection criteria of  \cite{mhamdi2024observational}, the models with \( 0 < |\Delta\text{AIC}| < 2 \) have substantial support, while those with \( 4 < |\Delta\text{AIC}| < 7 \) have considerably less support. In our case, for CC, $\Delta \text{AIC}=5.25<7$ and hence a moderate support is there, in light with \cite{mhamdi2024observational}. If we look at the $\chi^2_{min}$ for $\Lambda$CDM and our $f(Q)$ model as presented in Table \ref{T1} and \ref{table:01} respectively, we observe considerable closeness between the respective values. Nonetheless, the exponential $f(Q)$ model's ability to yield consistent cosmological parameter estimates and its moderate $\Delta$AIC values suggest that it remains an alternative, especially in light of its theoretical motivation to unify early and late-time cosmic acceleration.}

\begin{table}[hbt]
\renewcommand\arraystretch{1.5}
\centering 
\begin{tabular}{||c|c|c|c|c|c||} 
\hline\hline 
~~~Dataset~~~&~~~~~~~ $H_{0}$ ~~~~~~~& ~~~~~~~$\Omega_{m_0}$~~~~~~~ & ~~~~~~~$\chi^2_{min}$~~~~~~&~~~~~$AIC$~~~~~~&~~~~~$BIC$~~~~~\\ [0.5ex]  
\hline\hline
CC & $67.8\pm 1.8$ &   $0.314^{+0.060}_{-0.063}$ & $14.54$ & $18.54$& $21.47$ \\[0.5ex]
\hline
CC+Pantheon & $69.5\pm 1.5$ &  $0.301\pm0.046$ & $959.37$ & $963.37$& $973.34$\\[0.5ex]
\hline
CC + Pantheon$^+$ & $71.62^{+0.47}_{-0.49}$ &  $0.282^{+0.021}_{-0.024}$ & $1647.89$ & $1651.89$& $1662.81$\\[0.5ex]
\hline \hline 
\end{tabular}
\caption{\textcolor{black}{Constrained parameter values, $\chi^2_{min}$, AIC and BIC for the $\Lambda$CDM model are obtained using three observational datasets: CC, CC + Pantheon, and CC + Pantheon$^+$.}}
\label{T1} 
\end{table}

\begin{table}[hbt]
\renewcommand\arraystretch{1.5}
\centering 
\begin{tabular}{||c|c|c|c|c|c||} 
\hline\hline 
~~~Dataset~~~&~~~~~~~$\chi^2_{min}$~~~~~~&~~~~~$AIC$~~~~~~&~~~~~$BIC$~~~~~&~~~~~$\Delta_{AIC}$~~~~~&~~~~~$\Delta_{BIC}$~~~~\\ [0.5ex]  
\hline\hline
CC & $16.049$ & $24.05$ & $29.91$ & $5.51$  & $8.44$  \\[0.5ex]
\hline
CC+Pantheon & $967.502$ & $975.50$ & $995.44$ & $12.13$ & $22.1$ \\[0.5ex]
\hline
CC + Pantheon$^+$ & $1658.814$ & $1666.81$ & $1688.64$ &  $14.92$ & $25.83$  \\[0.5ex]
\hline \hline 
\end{tabular}
\caption{\textcolor{black}{$\chi^2_{min}$, AIC, BIC for the observationally-constrained $f(Q)$ gravity model and $\Delta$AIC, $\Delta$BIC (differences in comparison with the $\Lambda$CDM model) using three observational datasets: CC, CC + Pantheon, and CC + Pantheon$^+$.}}
\label{table:01} 
\end{table}

\subsubsection*{\textbf{Reconstruction of the Distance Modulus $\mu(z)$ vs. Redshift $z$ from CC, CC+Pantheon, and CC+Pantheon$^+$ Data: Comparison with $\Lambda$CDM and Pantheon$^+$ Observations}}

\begin{figure}[htbp]
\centering
\includegraphics[width=1.1\textwidth]{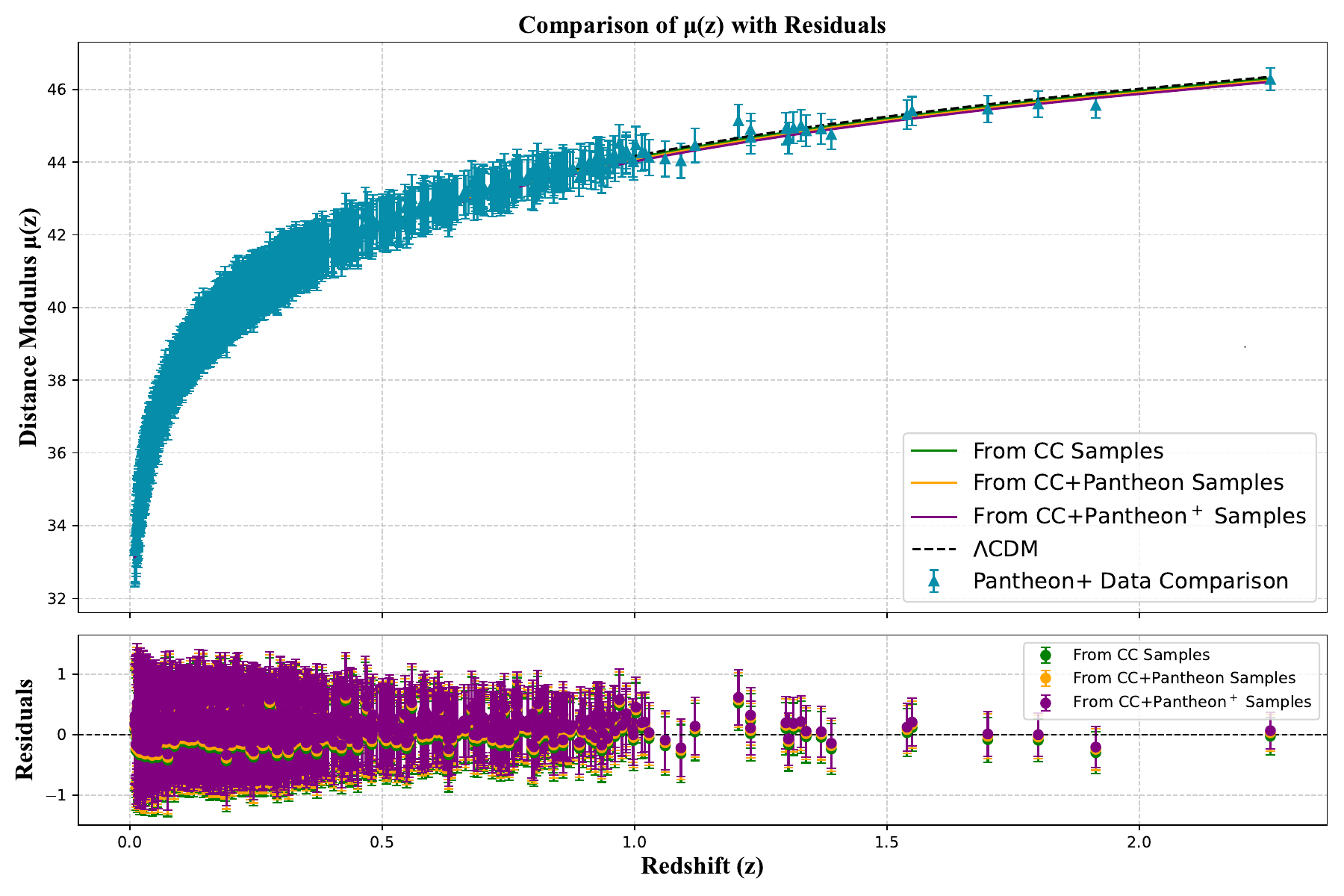}
\caption{Comparison of observed distance modulus \(\mu(z)\) from CC, CC+Pantheon, and CC+Pantheon$^+$ with predictions from the \(\Lambda\)CDM model. The upper panel shows the distance modulus versus redshift; the lower panel displays residuals, highlighting deviations from the observation. Error bars indicate standard uncertainties.}\label{fig3}
\end{figure}

In this subsection, we would illustrate through Figure~\ref{fig3} (Upper Panel) a comparison of the distance modulus $\mu(z)$ at different redshifts for three sets of parameters estimated from various combinations of 1701 Pantheon$^+$ data point. The curves, as derived from CC data alone (green), CC and Pantheon (orange), and CC and Pantheon$^+$ (magenta) are in consistency with the standard $\Lambda$CDM prediction (dashed black line), as well as with the observational data from the Pantheon$^+$ Supernovae compilation (teal blue error bars). This consistency helps us to interpret that, as we consider the CC dataset alone, we obtain statistically indistinguishable distance modulus predictions from those inferred when we use Supernovae data. This affirms the strength of CC measurements as an independent cosmological probe. The minor differences between the three colored curves imply that the inclusion of the Pantheon or Pantheon$^+$ samples marginally brings any change to the deduced expansion history. We infer from this that the mild change might be due to the high constraining power and accuracy of the present CC data. Furthermore, residuals between all models and the Pantheon$^+$ data are still above or below the error limits across the entire redshift range, including at $z \gtrsim 1.5$ where observational errors are larger. This pattern is consistent with internal consistency between various cosmological probes and supports the reconstructed $\mu(z)$ relations under minimal as well as extended data assumptions. The consistency of these model forecasts with the observational data set confirms the standard cosmological model and yields no substantial evidence for deviation from $\Lambda$CDM at the current error levels. More broadly, the plot depicts the validation of both observational techniques and theoretical model as developed in the present study.

We examined the residuals of the distance modulus $\mu(z)$ in Figure~\ref{fig3} (Lower Panel) for the 1701-point Pantheon$^+$ Supernovae dataset to assess the fidelity of our model in replicating the observed luminosity-distance relationship. The residuals are defined by the expression $\Delta \mu(z_i) = \mu_{\text{obs}}(z_i) - \mu_{\text{model}}(z_i)$, where $\mu_{\text{obs}}$ are the observational values from the Pantheon$^+$ catalogue, and $\mu_{\text{model}}$ are the theoretical predictions from our cosmological model for three sets of parameters derived respectively from CC only, CC combined with Pantheon, and CC combined with Pantheon$^+$. Unlike the residuals of $H(z)$, which directly track the expansion rate, the residuals in $\mu(z)$ reflect cumulative deviations in the integrated distance relation, making them more sensitive to subtle discrepancies in the background dynamics over redshift. Our residual analysis reveals that the model based on CC + Pantheon$^+$ parameters, along with other theoretical predictions from datasets, provides the closest alignment to the data, with residuals clustered tightly around zero and minimal systematic deviation, indicating that it better captures the luminosity-distance behavior across the redshift range than the other two parameter sets.

\subsection{Evolution of the Total Equation of State, Deceleration Parameter, and Fractional Energy Densities in \textcolor{black}{observationally-constrained} $f(Q)$ Gravity} 
The total EoS \cite{koussour2023thermodynamical} and deceleration parameter \cite{yadav2024reconstructing} in terms of $H(z)$ and redshift $z$, respectively, are given by:
\begin{equation}\label{E1}
\omega_{total}=-1+\frac{2(1+z)}{3H(z)}\frac{dH(z)}{dz},
\end{equation}
\begin{equation}\label{E2}
q=-1+\frac{(1+z)}{H(z)}\frac{dH(z)}{dz}.
\end{equation}
Hence, we can re-express the total EoS and the deceleration parameter respectively using Eqs.~\eqref{Eq:13},~\eqref{Eq:14} and \eqref{Eq:16}, as
\begin{equation}\label{Eq:19a}
    \omega_{total}=\frac{\eta_1  (8+\eta_2  \zeta (8+3 \eta_2  \zeta))-2 e^{\eta_2  \zeta} \left(4 \eta_1 +\zeta^2 (-1+\mu)\right)}{3 \left(2 e^{\eta_2  \zeta}-\eta_1
 \eta_2 ^2\right) \zeta^2},
\end{equation}
\begin{equation}\label{Eq:19b}
  q=  \frac{\eta_1  (2+\eta_2  \zeta)^2-e^{\eta_2  \zeta} \left(4 \eta_1 +\zeta^2 (-2+\mu)\right)}{\left(2 e^{\eta_2  \zeta}-\eta_1  \eta_2 ^2\right)
\zeta^2}.
\end{equation}
The \( \omega_{\text{total}} \) expression Eq.\eqref{Eq:19a} is a complicated nonlinear function of the parameters \( \eta_1 \), \( \eta_2 \), \( \zeta \), and \( \mu \). The numerator contains polynomial terms in \( \zeta \) and \( \eta_2 \), and a single exponential term \( e^{\eta_2 \zeta} \). In the subsequent phase, we would analyze the evolutionary behavior of \( \omega_{\text{total}} \). The constant and linear terms in the numerator and the \( \zeta^2 \) term in the denominator dominate the leading behavior in the early-time limit \( \zeta \to 0 \), assuming regular behavior of the exponential and polynomial terms. Extending \( e^{\eta_2 \zeta} \approx 1 + \eta_2 \zeta \), we find that \( \omega_{\text{total}} \to \infty \), which suggests a potential divergence or singular behavior at early times, though it depends on the values of \( \eta_1 \), \( \eta_2 \), and \( \mu \). We find that the exponential term dominates the behavior of both the numerator and the denominator when taking into account the late-time limit \( \zeta \to \infty \). Notably, if \( \mu = 1 \), the late-time universe behaves like a de Sitter phase with \( \omega_{\text{total}} \to 0 \), whereas \( \mu < 1 \) leads to \( \omega_{\text{total}} < 0 \), mimicking dark energy-like behavior.

In the early-time limit \( \zeta \to 0 \), using the expansion \( e^{\eta_2 \zeta} \approx 1 + \eta_2 \zeta \), the deceleration parameter \( q \) in Equation~\eqref{Eq:19b} shows divergent behavior due to the \( \zeta^2 \) term in the denominator. Hence, a a possible decelerated phase at early times is apparent, although it depends on the values of \( \eta_1 \), \( \eta_2 \), and \( \mu \). In the late-time limit \( \zeta \to \infty \), where the exponential term dominates, \( q \) asymptotically approaches a constant. For \( \mu < 2 \), the deceleration parameter becomes negative, implying an accelerating universe, while \( \mu = 2 \) corresponds to the transition with \( q = 0 \) as the transition from deceleration to acceleration requires \(q=0\) at some stage.
\begin{figure}[htbp]
\centering
\includegraphics[width=1.1\textwidth]{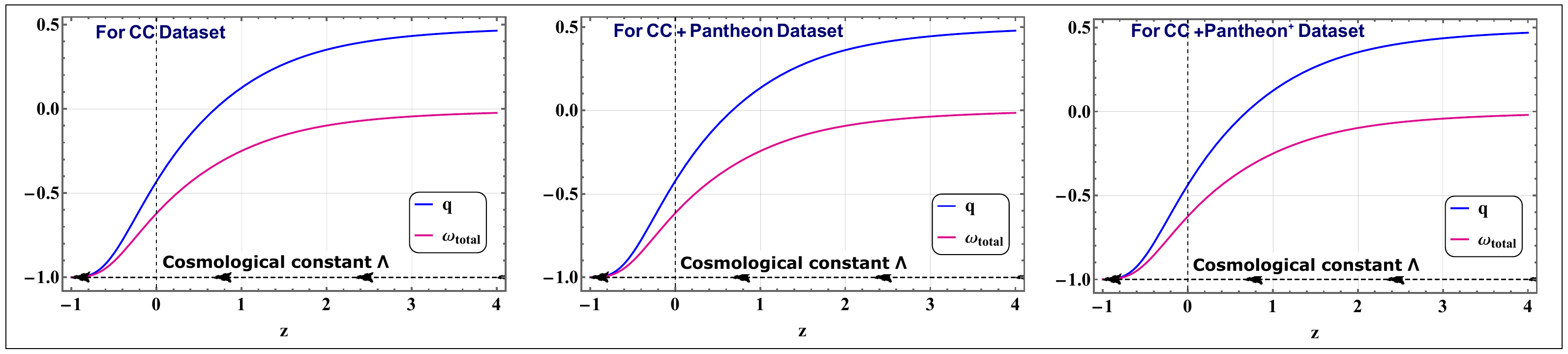}
\caption{Evolution of the $\omega_{\text{total}}$ and $q$ with redshift $z$, constrained using CC, CC+Pantheon, and CC+Pantheon$^+$ datasets.\label{fig4}}
\end{figure}

Figure~\ref{fig4} demonstrates the evolution of the total EoS parameter $\omega_{\text{total}}$ and the deceleration parameter $q$ as functions of redshift $z$, constrained using three observational datasets: CC alone, CC+Pantheon, and CC+Pantheon$^+$. In all cases, the present-day values of $\omega_{\text{total}} \approx -1$ and $q < 0$ are consistent with late-time accelerating universe. As we approach from lower to higher redshifts, both parameters transition smoothly toward values characteristic of the matter-dominated era, with $q > 0$ and $\omega_{\text{total}} \to 0$. The consistency across datasets further establishes the viability of the \textcolor{black}{observationally-constrained} $f(Q)$ model for evolution of the universe. 

\begin{figure}[htbp]
\centering
\includegraphics[width=1.1\textwidth]{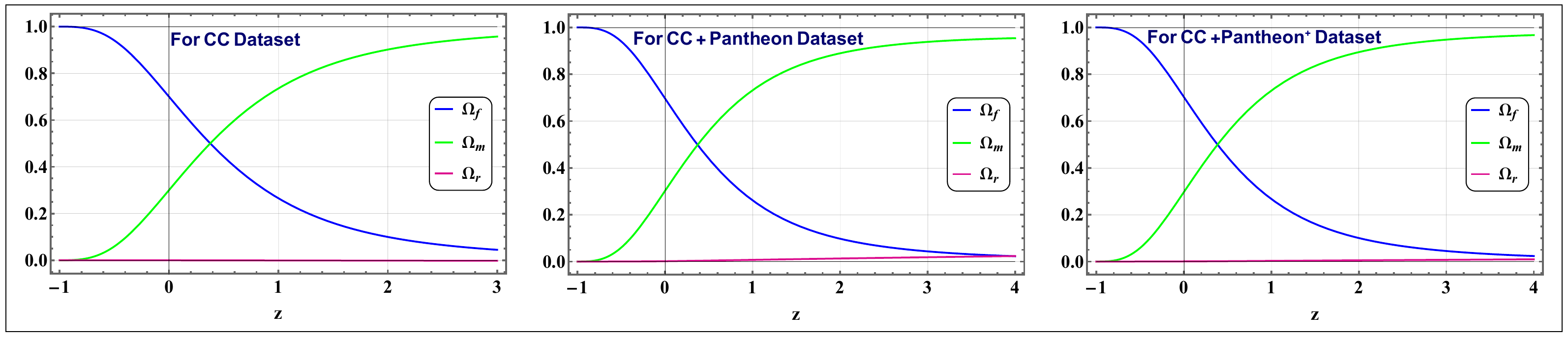}
\caption{Redshift evolution of the fractional energy densities in the \textcolor{black}{observationally-constrained} $f(Q)$ gravity model, constrained using CC, CC combined with the Pantheon Type Ia Supernovae sample, and CC combined with Pantheon$^+$. The plots depict a transition from a matter-dominated epoch at higher redshifts to a dark energy-dominated phase at lower redshifts, highlighting the shift in the primary driver of cosmic expansion.\label{fig5}}
\end{figure}

In Figure~\ref{fig5}, three observational datasets were used to constrain the redshift evolution of the fractional energy densities \cite{bhagat2025tracing} in the \textcolor{black}{observationally-constrained} gravity model: CC, CC combined with the Pantheon Type Ia Supernovae sample, and CC combined with Pantheon$^+$. The plots illustrate the change in the main force behind cosmic expansion, showing a shift from a matter-dominated epoch at higher redshifts to a dark energy-dominated phase at lower redshifts. This transition underscores the shift in the primary driver of cosmic expansion from matter to dark energy. The present-day values of the dark energy density parameter, \( \Omega_{\mathrm{de}} \), are approximately in $[0.69,0.8]$ across the datasets. These findings indicate that the universe is currently in a phase dominated by dark energy, consistent with the observed accelerated expansion. The evolution profiles also suggest that dark energy will continue to dominate the cosmic dynamics in the foreseeable future.

\subsection{Probing Cosmic Dynamics in \textcolor{black}{observationally-constrained} $f(Q)$ Gravity through Statefinder and $Om(z)$ Diagnostics}

\begin{figure}[htbp]
\centering
\includegraphics[width=1.1\textwidth]{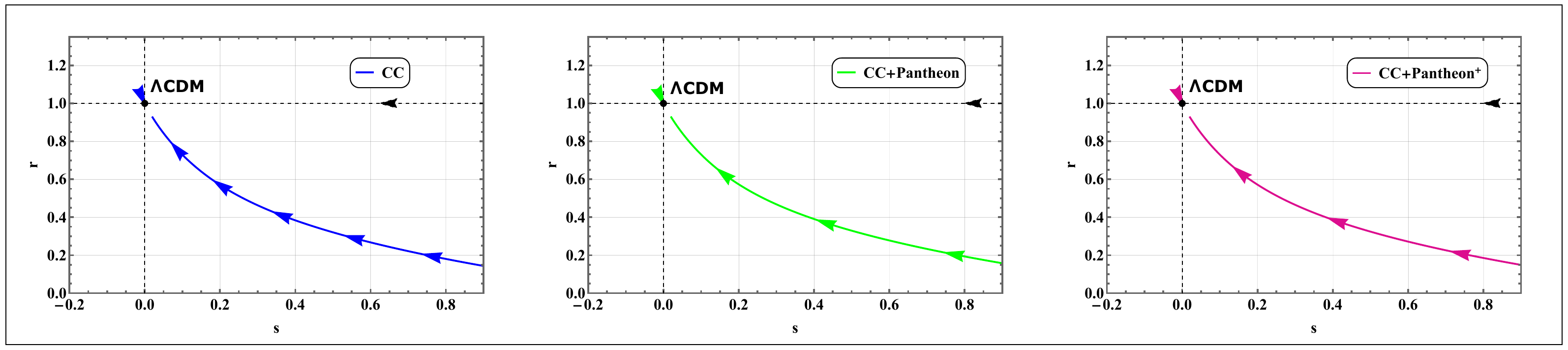}
\caption{Evolution of the $r$–$s$ Statefinder trajectories for the \textcolor{black}{observationally-constrained} $f(Q)$ gravity model constrained using CC, CC+Pantheon, and CC+Pantheon$^+$ datasets.\label{fig6}}
\end{figure}
To gain a better understanding of different cosmological models, particularly in relation to dark energy, the Statefinder diagnostic has been introduced as a valuable geometric tool. This diagnostic is derived from higher derivatives of the scale factor, allowing for a model-independent method to compare theoretical predictions with observational data. The cosmological diagnostic pair \(\{r, s\}\) \cite{Sahni_2003_77,alam2003exploring} is directly linked to the scale factor \(a\) and, therefore, to the metric that describes space-time. For this reason, it is considered ``geometric." The parameters \(r\) and \(s\) are defined in terms of the Hubble parameter \(H\) and its time derivatives, respectively, as 
\begin{equation}
r=1+3\frac{\dot{H}}{H^{2}}+\frac{\ddot{H}}{H^{3}},
    \label{r1}
\end{equation}
\begin{equation}
s=-\frac{3H\dot{H}+\ddot{H}}{3H(2\dot{H}+3H^{2})}.
    \label{r2}
\end{equation}
We can also re-express the Statefinder parameter $r$ in terms of \(H(z)\) and redshift $z$, and the Statefinder $s$ in terms of $r$ and $q$, respectively, as
\begin{equation}
r(z) = 1 - 2(1+z) \frac{H'(z)}{H(z)} + (1+z)^2 \left[ \frac{H''(z)}{H(z)} + \left( \frac{H'(z)}{H(z)} \right)^2 \right],
\label{r_z}
\end{equation}
\begin{equation}
s=\frac{r(z) - 1}{3\left(q(z) - \frac{1}{2}\right)},
\label{s_z}
\end{equation}
where $H'=\frac{dH}{dz}$ and $H''=\frac{d^2H}{dz^2}$. Table~\ref{rs} presents the Statefinder parameters \((r, s)\) for various cosmological models, illustrating how different dark energy scenarios occupy distinct regions in the \((r, s)\)-plane.

\begin{table}[h!]
\centering
\renewcommand{\arraystretch}{1.4} 
\begin{tabularx}{\textwidth}{|>{\centering\arraybackslash}X
                              |>{\centering\arraybackslash}X
                              |>{\centering\arraybackslash}X|}
\hline \hline
\textbf{Corresponding Model} & \textbf{\(r\)} & \textbf{\(s\)} \\
\hline \hline
\(\Lambda\)CDM model & \(1\) & \(0\) \\
\hline
Chaplygin Gas (CG) model & \(>1\) & \(<0\) \\
\hline
Standard Cold Dark Matter (SCDM) model & \(1\) & \(1\) \\
\hline
Quintessence region & \(<1\) & \(>0\) \\
\hline
Holographic Dark Energy (HDE) model & \(1\) & \(\frac{2}{3}\) \\
\hline \hline
\end{tabularx}
\caption{Statefinder parameters \((r, s)\) for different cosmological models.}
\label{rs}
\end{table}

Figure~\ref{fig6} depicts the evolution of the $r$–$s$ Statefinder trajectory for the \textcolor{black}{observationally-constrained} $f(Q)$ gravity model. The model is constrained using CC, CC combined with Pantheon, and CC combined with Pantheon$^+$.  It is well known that the Statefinder parameters $\{r, s\}$ provide us with a powerful tool to discriminate between models of cosmology by examining higher-order derivatives of the scale factor. The fixed point $\{r = 1, s = 0\}$ corresponds to the standard $\Lambda$CDM model. As evident from all three panels, the trajectories begin from higher values of $s$ and lower $r$, eventually converging toward the $\Lambda$CDM point. This convergence indicates that the \textcolor{black}{observationally-constrained} $f(Q)$ model asymptotically behaves like the $\Lambda$CDM model at late times, thereby ensuring consistency with observational cosmology. The direction of arrows along each curve depicts the evolution of the universe from the early to the late stage. Across all three datasets, we observe that the evolution resembles a consistent transition from an early-time decelerated phase (lower $r$, higher $s$) to a late-time accelerated phase that is compatible with accelerated expansion of the dark-energy dominated universe. Thus, through Statefinder analysis, we understand that the \textcolor{black}{observationally-constrained} $f(Q)$ model is consistent with observation and holds dynamic viability. One specific observation is that despite blending CC with other datasets, we have almost similar steepness of the trajectory from early to late time.
\begin{figure}[htbp]
\centering
\includegraphics[width=1.1\textwidth]{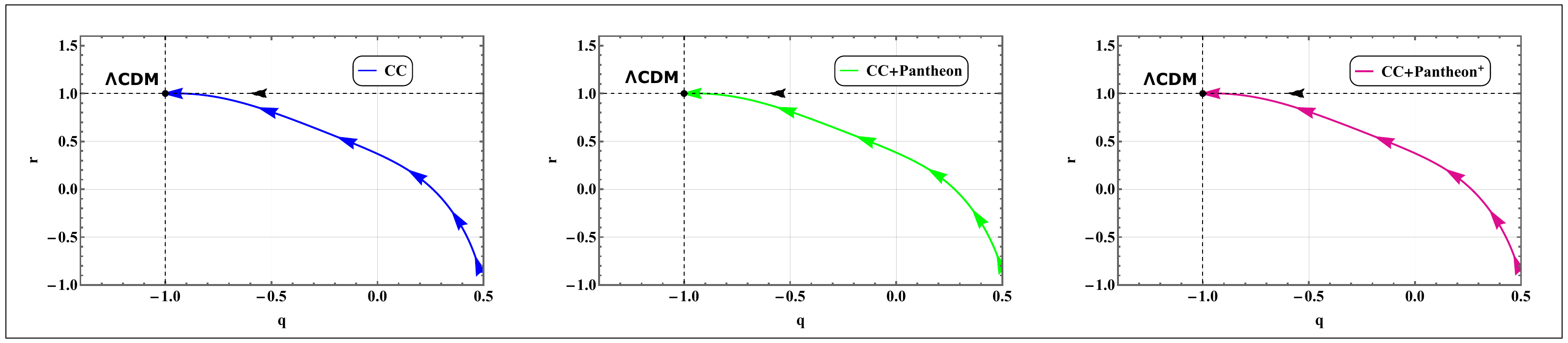}
\caption{$r$–$q$ trajectories for the \textcolor{black}{observationally-constrained} $f(Q)$ gravity model constrained using CC, CC+Pantheon, and CC+Pantheon$^+$ datasets.\label{fig7}}
\end{figure}

Figure~\ref{fig7} represents the $r$–$q$ trajectories for three different observational constraints: CC, CC+Pantheon, and CC+Pantheon$^+$. The deceleration parameter $q$ transits from positive $q$ to negative $q$ signifying evolution of the universe from decelerated to accelerated phase. In all three panels, the trajectories begin from a region with $q>0$ and low value of Statefinder parameter $r$, that are corresponding to a decelerated early universe. Subsequently, the trajectories evolve toward the fixed point $\{r = 1, q = -1\}$, which is the fixed point pertaining to $\Lambda$CDM model. This convergence helps us interpret that the \textcolor{black}{observationally-constrained} $f(Q)$ model favors late-time acceleration alongside dynamically getting hold of $\Lambda$CDM in the far future. Like $\{r-s\}$, the $\{r-q\}$ also shows similar trajectory pattern across the three panels.

\begin{figure}[htbp]
\centering
\includegraphics[width=0.9\textwidth]{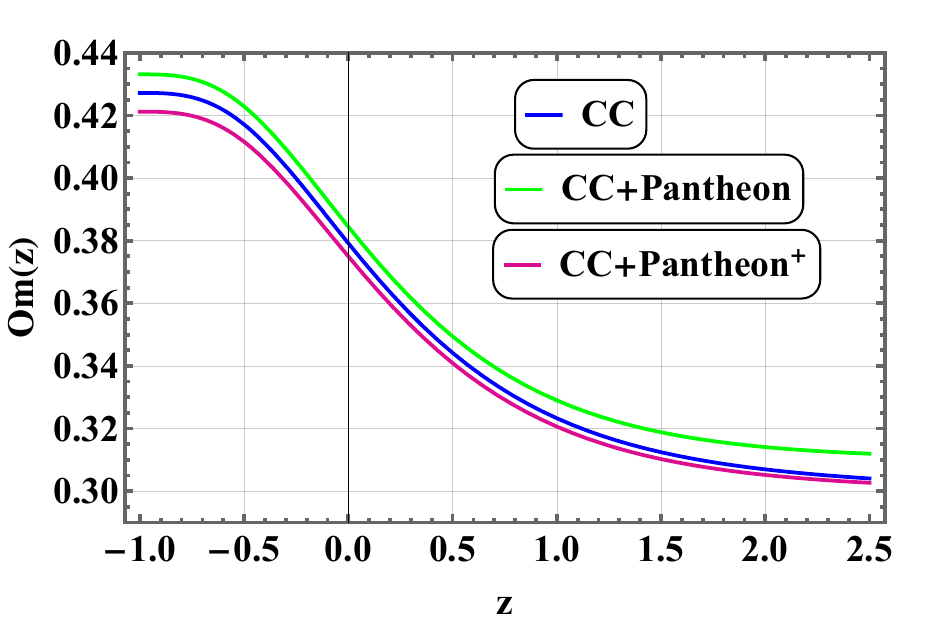}
\caption{The $Om(z)$ trajectories for different datasets show mild redshift dependence, with values clustering around $0.3$–$0.4$, broadly consistent with current estimates of $\Omega_{m_0}$.\label{fig8}}
\end{figure}
The $Om(z)$ diagnostic Sahni et al. (2008) \cite{sahni2008two}, is a diagnostic for investigating the cosmic expansion history in a model-independent way. It is given by:
\begin{equation}
Om(z) \equiv \frac{(H(z)/H_0)^2 - 1}{(1+z)^3 - 1}
\label{eq:om_z_def}
\end{equation}
where $H(z)$ is the Hubble parameter at redshift $z$, and $H_0$ is the Hubble constant. The $Om(z)$ diagnostic offers a model-independent tool to test the validity of $\Lambda$CDM by probing the total matter density parameter $\Omega_{m_0}$ without requiring a specific form of the dark energy equation of state $w(z)$. In a spatially flat $\Lambda$CDM model, $Om(z)$ remains constant at $\Omega_{m_0}$; thus, any deviation from constancy can indicate evolving dark energy or departures from spatial flatness. The evolution of $Om(z)$ for three combinations of observational data is shown in Figure~\ref{fig8}. The robustness of $H(z)$ constraints obtained from these datasets is further supported by the close alignment of the CC-only (blue line) and CC+Pantheon$^+$ (magenta line) curves, which show mutual consistency. At higher redshifts, however, the CC+Pantheon (green line) trajectory shows a discernible divergence, suggesting potential systematic tensions or discrepancies between the extended Pantheon$^+$ datasets and the canonical Pantheon. All trajectories exhibit a mild redshift dependence, with $Om(z)$ values clustering around $0.3$–$0.4$. This is generally consistent with current estimates of $\Omega_{m_0}$. The general downward trend across redshift is qualitatively consistent with the transition from a decelerated, matter-dominated phase to an accelerated, dark energy-dominated epoch. These results show the utility of the $Om(z)$ diagnostic in analyzing the complementarity and consistency of cosmological data.

\subsection{Age Evolution of the Universe in \textcolor{black}{observationally-constrained} $f(Q)$ Gravity Framework}

\begin{figure}[htbp]
\centering
\includegraphics[width=0.9\textwidth]{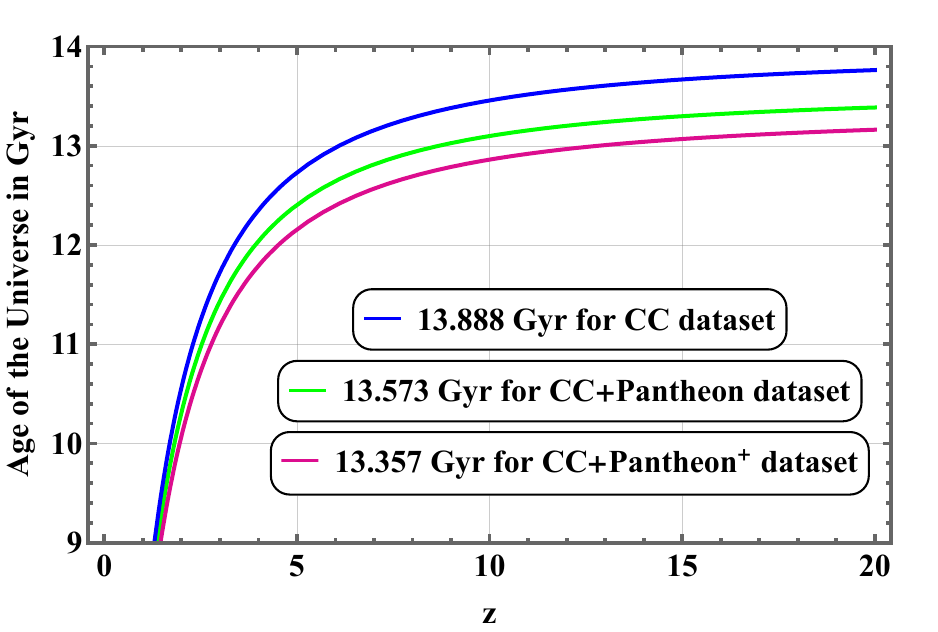}
\caption{Age evolution of the universe for the CC, CC+Pantheon, and CC+Pantheon$^+$ datasets as a function of redshift $z$. The current age estimates for the datasets mentioned are $13.888$, $13.573$, and $13.357$ Gyr, respectively. The viability of the \textcolor{black}{observationally-constrained} $f(Q)$ model is supported by its good alignment with Planck bounds \cite{aghanim2020planck}.\label{fig9}}
\end{figure}
The age of the universe $t_0$ is estimated by 
\begin{equation}
    t_0 = \int_0^{\infty} \frac{dz}{(1+z)H(z)}.
\end{equation}
This relation connects the expansion of the universe to its overall age, thus providing a fundamental consistency check for cosmological models.
Figure~\ref{fig9} shows the age evolution of the universe as a function of redshift $z$ for three different observational constraints: CC, CC+Pantheon, and CC+Pantheon$^+$. The CC dataset's current age is $13.888$ Gyr, the CC+Pantheon dataset's is $13.573$ Gyr, and the CC+Pantheon$^+$ dataset's is $13.357$ Gyr. These findings show that, with minor deviations brought about by the inclusion of Supernovae data, the estimate of cosmic age remains constant across all dataset combinations. The \textcolor{black}{observationally-constrained} $f(Q)$ model is consistent in explaining the evolution of the universe since the obtained values remain consistent with the current bounds from Planck \cite{aghanim2020planck} and other observational probes \cite{wong2020h0licow}.

\subsection{Energy Conditions and Effective Energy Density Evolution}
This subsection will discuss various energy conditions, which are combinations of energy density and pressure. The literature identifies four main types of energy conditions: the weak energy condition (WEC), strong energy condition (SEC), null energy condition (NEC), and dominant energy condition (DEC). These energy conditions \cite{pradhan2024cosmological,yadav2024reconstructing} are presented in Table~\ref{EC}.
\begin{table}[h!]
\centering
\renewcommand{\arraystretch}{1.4}
\begin{tabular}{|c|c|c|}
\hline
\textbf{Condition} & \textbf{Inequalities} & \textbf{In Terms of \( \omega = p/\rho \)} \\
\hline
NEC & \( \rho + p \geq 0 \) & \( w \geq -1 \) (if \( \rho > 0 \)) \\
\hline
WEC & \( \rho \geq 0,\ \rho + p \geq 0 \) & \( w \geq -1 \) \\
\hline
SEC & \( \rho + p \geq 0,\ \rho + 3p \geq 0 \) & \( w \geq -\frac{1}{3} \) \\
\hline
DEC & \( \rho \geq 0,\ \rho - p \geq 0 \) & \( w \leq 1 \) \\
\hline
\end{tabular}
\caption{Standard energy conditions expressed in terms of energy inequalities and the equation of state parameter \( \omega = p/\rho \).}
\label{EC}
\end{table}

\begin{figure}[htbp]
\centering
\includegraphics[width=1.1\textwidth]{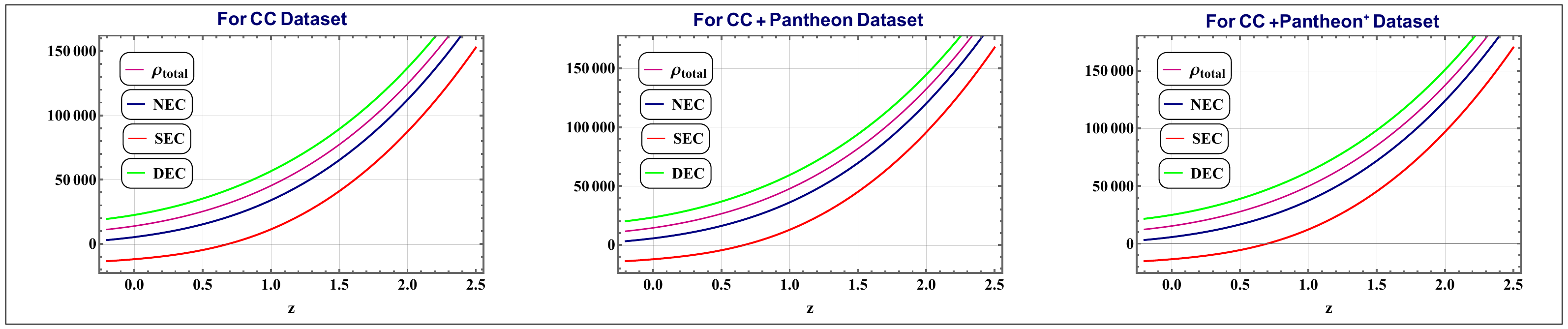}
\caption{The behavior of the Null Energy Condition (NEC), Strong Energy Condition (SEC), and Dominant Energy Condition (DEC), along with the total effective energy density $\rho_{\text{total}}$, as functions of redshift $z$ for different observational constraints.\label{fig10}}
\end{figure}

The behavior of the Null Energy Condition (NEC), Strong Energy Condition (SEC), and Dominant Energy Condition (DEC) as well as the total effective energy density $\rho_{\text{total}}$ as functions of redshift $z$, for various combinations of observational constraints, is depicted in Figure~\ref{fig10}. The physical plausibility of the \textcolor{black}{observationally-constrained} $f(Q)$ gravity model is theoretically checked by these energy conditions, which are derived from the Raychaudhuri equation and Einstein's field equations. While SEC is violated at lower $z$, which is consistent with the presence of late-time cosmic acceleration, NEC and DEC are satisfied throughout the redshift range in all three panels. The model's feasibility is further supported by the evolution of $\rho_{\text{total}}$, which shows monotonic growth with increasing redshift.

\section{Concluding remarks}

Modified gravity theories have become pivotal in addressing the limitations of General Relativity (GR) in explaining cosmic acceleration without invoking dark energy \citep{Nojiri2011, Clifton2012}. Among these, the $f(Q)$ gravity framework, rooted in symmetric teleparallelism, has recently emerged as a promising alternative where gravity is described via the non-metricity scalar $Q$, and not curvature or torsion \cite{Jimenez2018, BeltranJimenez2019}. In the work presented in this paper, we have used a modified $f(Q)$ theory to investigate the cosmological consequences of symmetric teleparallel gravity. After reviewing the general framework of $f(Q)$ gravity, we built a particular exponential model with two free parameters. In order to investigate both early-time inflationary and late-time accelerated expansion scenarios, this model was created to offer a smooth departure from General Relativity. We obtained constraints on the model parameters by using Markov Chain Monte Carlo (MCMC) methods to confront the model with a variety of current cosmological observations. We then used important cosmological diagnostics, such as the deceleration parameter, equation of state (EoS) parameter, energy density parameters, Statefinder and Om diagnostics, and others, to analyze the dynamical behavior of the model.

In the context of symmetric teleparallel gravity, we have constructed and examined an exponential $f(Q)$ gravity model in this paper. By adding a smooth deviation parameterized by two free parameters, $\eta_1$ and $\eta_2$, which regulate the magnitude and strength of the modification, this construction goes beyond the conventional General Relativity paradigm. The cosmological viability of the proposed exponential $f(Q)$ model, defined by the functional form $f(Q) = Q + \eta_1 Q_0\left(1 - e^{-\eta_2 \sqrt{Q/Q_0}}\right)$ in Eq.~\eqref{Eq:013}, has been thoroughly examined, demonstrating that the model admits a smooth transition between early and late time cosmic evolution through appropriate choices of the free parameters $\eta_1$ and $\eta_2$. We have derived expressions for the effective density parameters in Eq.~\eqref{Eq:16} and the deceleration parameter in Eq.~\eqref{Eq:19b} by introducing the dimensionless variables defined in Eq.~\eqref{Eq:14} and taking into account the standard energy content of a universe filled with dust and radiation as given in Eq.~\eqref{Eq:15}. These serve as the basis for studying the cosmological evolution in the case of exponential $f(Q)$ gravity. Additionally, a thorough numerical analysis of the model's behavior over cosmic time is made possible by the dynamical system defined by the coupled differential equations \eqref{Eq:20} and \eqref{Eq:21}. This formalism, which smoothly interpolates between radiation domination, matter domination, and the current accelerated expansion, captures the model's capacity to explain a plausible evolution of the universe.

In the next phase of the study, using recent observational datasets, we conducted a thorough parameter estimation of the exponential $f(Q)$ gravity model in Section~4. In particular, we used the CC dataset \cite{moresco20166}, the Pantheon Supernovae compilation \cite{carr2022pantheon}, and the Pantheon$^+$ sample \cite{brout2022pantheonplus} to constrain the model parameters $\eta_1$, $\eta_2$, $H_0$, and $\Omega_{m_0}$ using Markov Chain Monte Carlo (MCMC) techniques \cite{Lewis2002, foreman2013emcee}. According to the results, the parameter space supports a late-time acceleration scenario, and the adopted $f(Q)$ model is still consistent with current cosmological observations \cite{Lazkoz2021, Jimenez2020}. The model's feasibility within observational bounds is further supported by the agreement between the best-fit values and the $\Lambda$CDM baseline.

\textcolor{black}{Inspired by the foundational work of \cite{mhamdi2024constraints}, the study reported in the present paper explores $f(Q)$ gravity. Although conceptually inspired, the current study has implemented a somewhat different methodology. Here, we propose a specific exponential functional form for $f(Q)$ and utilize MCMC techniques to constrain its parameters against cosmic chronometers (CC), Pantheon, and Pantheon$^+$ Supernovae datasets. This approach is in contrast with \cite{mhamdi2024constraints}, which explored different $f(Q)$ forms and focused on parameter estimation using  dataset combinations of Pantheon+H(z) and Pantheon+H(z)+RSD. Apart from parameter constraints, our analysis further provides a comprehensive dynamical study that includes the evolution of the EoS parameter and the deceleration parameter, Statefinder and $Om(z)$ diagnostics, as well as a discussion on energy conditions and cosmological age estimates. This detailed exploration of cosmological diagnostics offers a more holistic understanding compared to previous studies, including \cite{mhamdi2024constraints}, which primarily emphasized parameter constraints.}

The corner plot in Figure~\ref{fig1} illustrates how effectively several datasets can be used to tighten the parameter constraints of the exponential $f(Q)$ gravity model. The addition of Pantheon and Pantheon$^+$ Supernovae data significantly narrows the posterior distributions, especially for $H_0$ and $\Omega_{m_0}$, indicating increased precision \cite{scolnic2018complete,brout2022pantheonplus}. The model's stability and observational feasibility are indicated by the contours' consistent overlap across all datasets. These findings show that the model retains a high degree of agreement with available cosmological data and that parameter degeneracies are significantly reduced by the inclusion of complementary datasets \cite{jimenez2002constraining,moresco2012improved}. To further consolidate the results depicted in Figure~\ref{fig1}, we have presented correlation matrices in Figure~\ref{fig01} in the form of a heat map to visualize the association among different parameters in $\Theta$. A strong anti-correlation has been observed between $H_0$ and $\Omega_{m_0}$ in CC, CC + Pantheon, and CC + Pantheon$^+$. However, the association of $H_0$ with $\eta_1$ and $\eta_2$ is low $(< 0.5)$ across the datasets and notably it is close to $0$ for $(H_0,\eta_1)$ in case of CC + Pantheon$^+$. A moderately strong anti-correlation has been observed between $\Omega_{m_0}$ and $\eta_1$ in all three cases, whereas the association is too weak $(\approx 0)$ for $(\Omega_{m_0},\eta_2)$, although little improvement appears in CC + Pantheon$^+$. The robustness and observational viability of the proposed \(f(Q)\) model are demonstrated by the consistent constraints on \(\Omega_{m_0}\), \(\eta_1\), and \(\eta_2\) across all datasets, as well as the increased precision and elevated central value of \(H_0\) upon including Pantheon data (see Fig.~\ref{fig02}). With the CC+Pantheon$^+$ combination showing slight departures from $\Lambda$CDM at higher redshifts ($z \gtrsim 1.5$), the comparative analysis of the reconstructed Hubble parameter $H(z)$ in Figure~\ref{fig2} highlights the value of combining complementary datasets and highlights the usefulness of multi-probe observations in improving cosmological predictions \cite{yang2021observational,lu2023observational}.  \textcolor{black}{Our analysis, as depicted in Figure~\ref{f1}, exhibits the robustness of $\Lambda$CDM and the effectiveness of data combination. While CC data are in alignment with Planck 2018 results, combining CC with Supernovae significantly tightens constraints on $H_0$ and $\Omega_{m_0}$, as shown in this Figure~\ref{f1}. Next, we implemented the AIC and BIC to statistically assess the exponential $f(Q)$ gravity model over $\Lambda$CDM. The $\Lambda$CDM model appears to be statistically preferred as indicated by lower AIC and BIC values, even though it produces consistent cosmological parameter estimates and shows moderate support via $\Delta$AIC (e.g., $\Delta\text{AIC}=5.25$ for CC, within the $5 < \Delta \mathrm{AIC} < 15$ range), as summarised in Table~\ref{T1} and \ref{table:01}. However, the similar $\chi^2_{\rm min}$ values of the exponential $f(Q)$ model and its theoretical justification for unifying cosmic acceleration indicate that it is still a feasible option.} As seen in Figure~\ref{fig3}, the reconstructed distance modulus $\mu(z)$ closely agrees with the $\Lambda$CDM prediction and Pantheon$^+$ data, demonstrating the robustness of CC data as a trustworthy stand-alone cosmological probe \cite{moresco2012improved}. The residual plots Figure~\ref{fig2} and \ref{fig3} (Lower Panel) for both $H(z)$ from 32 CC data points and $\mu(z)$ from 1701 Pantheon$^+$ Supernovae respectively reveal the level of agreement between theoretical predictions and observational data. The models constructed using only CC data, combined datasets CC + Pantheon and CC + Pantheon$^+$ demonstrate progressively smaller residuals, highlighting improved consistency. The CC + Pantheon$^+$ model, in particular, achieves the best overall fit in both Hubble and distance modulus observations. This indicates that integrating diverse datasets enhances parameter constraints and yields more reliable cosmological predictions across different redshift ranges. As shown in Figure~\ref{fig4}, the smooth transition of $\omega_{\text{total}}$ and $q$ from late-time acceleration to matter domination confirms that the \textcolor{black}{observationally-constrained} $f(Q)$ model consistently describes cosmic evolution across datasets \cite{nojiri2005modified}. These results underscore the viability of the \textcolor{black}{observationally-constrained} $f(Q)$ gravity model in capturing the universe's transition from matter to dark energy domination, aligning well with observational data and offering a compelling alternative framework to explain the current accelerated expansion \cite{jimenez2023reconstructing,koivisto2018metric}. Figure~\ref{fig5} demonstrates that, across all three observational datasets CC, CC combined with the Pantheon Type Ia Supernovae sample, and CC combined with Pantheon$^+$, the \textcolor{black}{observationally-constrained} $f(Q)$ gravity model exhibits a transition from a matter-dominated epoch at higher redshifts to a dark energy-dominated phase at lower redshifts, with present-day values of the dark energy density parameter \( \Omega_{\mathrm{de}} \) lying within the range $[0.69, 0.8]$, indicating the current dominance of dark energy in the universe's accelerated expansion.

A cosmological evolution consistent with observational data is shown by the \textcolor{black}{fitted} \( f(Q) \) gravity model, constrained by CC \cite{moresco2012improved}, CC combined with the Pantheon Supernovae Type Ia sample \cite{scolnic2018complete}, and CC combined with Pantheon$^+$ datasets \cite{brout2022pantheonplus}. A shift from a decelerated, matter-dominated phase to an accelerated, dark energy-dominated epoch is shown by the Statefinder diagnostics (\( r \)–\( s \) and \( r \)–\( q \) trajectories), which converge towards the fixed point \( \{r = 1, s = 0\} \) characteristic of the $\Lambda$CDM model \cite{sahni2003Statefinder}. According to current estimates of the matter density parameter \( \Omega_{m_0} \) \cite{aghanim2020planck}, the $Om(z)$ diagnostic shows a mild redshift dependence, with values clustering around $0.3$–$0.4$. These results have been pictorially depicted in Figures~\ref{fig6}, \ref{fig7} and \ref{fig8}. In accordance with Planck bounds, the universe's age evolution, as shown in Figure~\ref{fig9}, produces current age estimates of roughly $13.35–13.89$ Gyr across the datasets. Figure~\ref{fig9} illustrates that the \textcolor{black}{observationally-constrained} $f(Q)$ gravity model yields consistent present-day age estimates of the universe, $13.888$ Gyr (CC), $13.573$ Gyr (CC+Pantheon), and $13.357$ Gyr (CC+Pantheon$^+$) demonstrating robustness across datasets and alignment with Planck constraints \cite{aghanim2020planck}, thereby affirming the model's viability in describing cosmic evolution.

 Moreover, the analysis of energy conditions shows that the Dominant Energy Condition (DEC) and Null Energy Condition (NEC) are satisfied across the redshift range, whereas the Strong Energy Condition (SEC) is violated at lower redshifts, which is consistent with late-time cosmic acceleration. Together, these findings support the \textcolor{black}{observationally-constrained} $f(Q)$ gravity model's feasibility in explaining the universe's accelerated expansion and its conformity to existing observational constraints (see Figure~\ref{fig10}).

In summary, the \textcolor{black}{observationally-constrained} exponential \( f(Q) \) gravity model demonstrates a smooth transition from early-time deceleration to late-time acceleration, aligning well with current observational data. This consistency underscores the model's viability as an alternative to \( \Lambda \)CDM in explaining the universe's accelerated expansion. In the context of symmetric teleparallel gravity, we developed and examined an exponential $f(Q)$ gravity model in this work. We used a variety of observational datasets, such as CC, Pantheon, and Pantheon$^+$ Type Ia Supernovae, to constrain the model parameters using Markov Chain Monte Carlo (MCMC) techniques. With parameter estimates demonstrating strong consistency across various data combinations, the analysis demonstrated that the model offers a solid description of the expansion history of the universe. Notably, the constraints on $H_0$ and $\Omega_{m_0}$ were considerably tightened by the addition of Pantheon and Pantheon$^+$ data, indicating the increased accuracy provided by these expanded datasets. The reconstructed distance modulus $\mu(z)$ was largely consistent with $\Lambda$CDM and observational data across all datasets, although slight deviations from the $\Lambda$CDM model were seen at higher redshifts for $H(z)$ with the most stringent constraints. Strong evidence for the feasibility of exponential $f(Q)$ gravity as a substitute cosmological model is presented in this work, which also emphasises the significance of multi-probe cosmological measurements for future improvement and offers a consistent account of the evolution of the universe. The implications of this model for early universe cosmology and its potential to resolve other urgent problems in contemporary cosmology will be the main topics of future research.

\textcolor{black}{While concluding, let us make some comments on the outcomes of the present study with respect to the existing literature. The fitting of the $f(Q)$ gravity model using recent CC, Pantheon, and Pantheon$^+$ datasets yields parameter constraints and cosmic evolution trajectories that closely align with observational benchmarks. The improved agreement in the posterior distributions of $H_0$ and $\Omega_{m_0}$ reinforces earlier claims of $f(Q)$ gravity's viability as an alternative to $\Lambda$CDM \cite{jimenez2023reconstructing,yang2021observational}. Our results, including the Statefinder and $Om(z)$ diagnostics, confirm a smooth transition from a decelerated matter-dominated era to an accelerated expansion phase, consistent with findings in modified gravity studies \cite{nojiri2005modified,koivisto2018metric}. Furthermore, the robustness of cosmic age estimates across datasets that are ranging between $13.35$ and $13.89$ Gyr supports compatibility with Planck constraints \cite{aghanim2020planck}. Most prior studies in $f(Q)$ gravity focus on simple polynomial or power-law modifications \cite{Mandour2023, Atayde2021, Frusciante2021}, often limited to specific cosmological epochs or narrow observational comparisons. These models generally offer limited flexibility in smoothly connecting the early and late universe dynamics. The present work introduces a novel exponential $f(Q)$ form: 
$ f(Q) = Q + \eta_1 Q_0\left(1 - e^{-\eta_2 \sqrt{Q/Q_0}}\right)$, which is specifically constructed to retain consistency with GR at low-energy limits while enabling significant modifications at high energies. This exponential structure is considered to facilitate a natural unification of early inflationary behavior and late-time acceleration. Our work is further distinguished by the comprehensive application of recent observational datasets, including CC, Pantheon, and Pantheon$^+$ Supernovae compilations within an MCMC framework. It may be noted in this context that earlier works such as \cite{Lazkoz2019} and \cite{Xu2021} employed fewer datasets or focused on background evolution only. On the contrary, the current study relies on multi-probe approach with the endeavour of obtaining tighter constraints on model parameters ($\eta_1$, $\eta_2$, $H_0$, $\Omega_{m_0}$) that might improve the robustness of model viability. Furthermore, our analysis of energy conditions offers physical insight into the model's adherence to classical gravitational bounds. In this context let us mention two notable works by \cite{Bajardi2020} and \cite{Hohmann2021}, where the works provided theoretical evaluations and our model satisfies the Null and Dominant Energy Conditions while violating the Strong Energy Condition at low redshifts, aligning with the behavior expected from accelerating expansion. The demonstrated consistency with multi-probe observations highlights the potential of the $f(Q)$ gravity and paves the way to further addressing tensions in $H_0$ and structure formation as a future study.}

\textcolor{black}{To sum up, the works presented in this paper are attempted to add some results to the expanding literature on $f(Q)$ gravity, a framework where non-metricity is the source of gravity. Recent studies have shown that under certain parameter regimes, a number of $f(Q)$ models, including power-law, exponential, and logarithmic models, can closely resemble the $\Lambda$CDM cosmological evolution~\cite{mhamdi2024constraints, Najera_2023_524}. Furthermore, observational analyses indicate that exponential $f(Q)$ models are statistically viable in the context of the standard $\Lambda$CDM model~\cite{mhamdi2024constraints} and power-law variants~\cite{mhamdi2024constraints}. Other models, suggested in~\cite{Anagnostopoulos_2021_822} and constrained in the current study, are compatible with observational and late-time acceleration datasets, sometimes surpassing $\Lambda$CDM in certain regimes. Overall, the current study is an attempt to establish the compatibility of exponential $f(Q)$ model with late-time acceleration and observational datasets. Collectively, thes outcomes focus on the viability of $f(Q)$ gravity as a viable alternative to the standard cosmological paradigm.} \textcolor{black}{While concluding, let us mention that the results presented in Figures~\ref{fig2} and \ref{fig3} show that the model's predictions for \( H(z) \) and \( \mu(z) \) are in close agreement with those of the \( \Lambda \)CDM model, with deviations emerging primarily at higher redshifts where observational uncertainties are larger. Additionally, the Statefinder analysis in Figure~\ref{fig6} indicates that the model trajectory approaches the \( \Lambda \)CDM fixed point. These findings suggest that the model is broadly consistent with \( \Lambda \)CDM, although statistical evidence do not establish supremacy of the proposed model over the standard cosmological model. As future study, we propose to experiment with other forms of $f(Q)$ in extensive manner to have further insight into it.}

\textcolor{black}{While \(\Lambda\)CDM remains statistically favored, the exponential \(f(Q)\) model developed in the rigorous study reported in this work successfully accommodates cosmic acceleration, remains compatible with multi-probe datasets, and provides a theoretically motivated alternative having its root in symmetric teleparallel gravity. The results of this work underscore the cosmological viability of the exponential \( f(Q) \) gravity model, affirm its agreement with independent observations across redshifts, and motivate further exploration of non-metricity-based gravities as plausible candidates for explaining the universe’s accelerated expansion.}

\subsection*{Acknowledgments}

\textcolor{black}{The insightful comments of the anonymous reviewer are gratefully acknowledged.} The authors also acknowledge the Inter-University Centre for Astronomy and Astrophysics (IUCAA), Pune, India, for providing library and computing facilities during their scientific visits in 2024 and 2025, which supported the completion of this work.

\subsection*{Declaration of generative AI and AI-assisted technologies in the writing process}

During the preparation of this work the author(s) used Grammarly and Quillbot in order to improve the language and correct the grammar. After using this tool/service, the author(s) reviewed and edited the content as needed and take full responsibility for the content of the publication.


\end{document}